%% file: paper.tex
\documentclass[usenatbib]{mn2e}

\usepackage{amsmath}
\usepackage{natbib}
\usepackage{epsfig}
\usepackage{txfonts}
\usepackage{pict2e}
\usepackage{color}

\newcommand{\nmax}{n_{\rm max}}
\newcommand{\bs}{\boldsymbol}

\newcommand{\WignerSixJ}[6]{{\left\{\begin{array}{ccc}
#1 & #2 & #3 \\
#4 & #5 & #6 \end{array} \right\}}}

\input{Befehle}

\usepackage{hyperref}
\usepackage{grffile}
\usepackage{graphics}
\usepackage{booktabs}

\title[Recombination Spectrum]
{{\tt CosmoSpec}: Fast and detailed computation of the cosmological recombination radiation from hydrogen and helium}

\author[Chluba \& Ali-Ha\"imoud]{Jens~Chluba$^{1, 2}$\thanks{E-mail:jchluba@ast.cam.ac.uk} and Yacine~Ali-Ha\"imoud$^{2}$\thanks{E-mail:yacine@jhu.edu}
\\
$^{1}$ Kavli Institute for Cosmology Cambridge, Madingley Road, Cambridge, CB3 0HA, UK. \\
$^{2}$ Department of Physics and Astronomy, Johns Hopkins University, 
3400 N. Charles St, Baltimore, MD 21218, USA\\
}

\begin{document}

\date{\vspace{-2mm}{Accepted 2015 --. Received 2015 October 14}}

\maketitle

\begin{abstract}
We present the first fast and detailed computation of the cosmological recombination radiation released during the hydrogen (redshift $z\simeq 1300$) and helium ($z\simeq 2500$ and $z\simeq6000$) recombination epochs, introducing the code {\tt CosmoSpec}. Our computations include important radiative transfer effects, $500$-shell bound-bound and free-bound emission for all three species, the effects of electron scattering and free-free absorption as well as interspecies ($\ion{He}{ii}\rightarrow \ion{He}{i}\rightarrow\ion{H}{i}$) photon feedback. The latter effect modifies the shape and amplitude of the recombination radiation and {\tt CosmoSpec} improves significantly over previous treatments of it. Utilizing effective multilevel atom and conductance approaches, one calculation takes only $\simeq 15$ seconds on a standard laptop as opposed to days for previous computations. This is an important step towards detailed forecasts and feasibility studies considering the detection of the cosmological recombination lines and what one may hope to learn from the $\simeq 6.1$ photons emitted per hydrogen atom in the three recombination eras. We briefly illustrate some of the parameter dependencies and discuss remaining uncertainties in particular related to collisional processes and the neutral helium atom model.
\end{abstract}


\section{Introduction}
\label{sec:Intro}

The energy spectrum of the cosmic microwave background (CMB) allows probing the thermal history of the Universe \citep{Zeldovich1969, Sunyaev1970mu, Burigana1991, Hu1993}, with several interesting standard and non-standard signals awaiting us \citep[for overview see][]{Chluba2011therm, Sunyaev2013, Tashiro2014}. One of the guaranteed signals is imprinted by the cosmological recombination process \citep{Zeldovich68, Peebles68, Dubrovich1975} at redshifts $z\simeq 10^3-10^4$. It appears as a spectral distortion of the CMB that today is visible at frequencies $\nu\simeq 0.1\,\GHz-2\,{\rm THz}$ \citep{Sunyaev2009}. 
The cosmological recombination radiation (CRR) constitutes a fundamental signal from the early Universe that can provide an alternative way to measure some of the key cosmological parameters \citep[e.g.,][]{Chluba2008T0, Sunyaev2009} and allow testing the physics of the cosmological recombination process. As challenging as the detection of the CRR may be \citep{Mayuri2015, Vince2015, Balashev2015}, it is a highly desirable target for future CMB studies \citep{Silk2014Sci}, and may be possible with an improved version of {\it PIXIE} \citep{Kogut2011} or {\it PRISM} \citep{PRISM2013WPII}.

Nowadays, the detailed recombination history (the free electron fraction as a function of redshift) can be computed with $\lesssim 0.1\%$ precision around decoupling ($z\simeq 10^3$) using {\tt CosmoRec} \citep{Chluba2010b} and {\tt HyRec} \citep{Yacine2010c}. This precision is required to ensure that our interpretation of CMB data from {\it Planck} gives unbiased results in particular for the spectral index of primordial fluctuation and the baryon density \citep{Jose2010, Shaw2011, Planck2015params}.

While in terms of standard (atomic) physics the recombination history is expected to be well understood, non-standard process, for instance, due to {\it decaying} or {\it annihilating particles} \citep[e.g.,][]{Peebles2000, Chen2004, Padmanabhan2005, Galli2009, Huetsi2009, Slatyer2009, Giesen2012} or {\it variations of fundamental constants} \citep[e.g.,][]{Kaplinghat1999, Battye2001, Rocha2003, Scoccola2009, Galli2009b} can modify the ionization history. These scenarios can be constrained with CMB temperature and polarization data \citep{Planck2015params, Planck2015var_alp}. It is, however, hard to discern between different sources of changes to the recombination history. Similarly, when allowing standard parameters such as the helium mass fraction, $Y_{\rm p}$, and effective number of neutrino species, $N_{\rm eff}$, to vary simultaneously, large degeneracies remain \citep{Planck2015params}. 
In all these cases, the CRR is expected to provide extra pieces of the puzzle, which will help to break degeneracies and directly probe the recombination dynamics \citep{Chluba2008T0, Sunyaev2009, Chluba2010a}. In addition, the reaction of the plasma in the pre-recombination era can give hints about early energy release \citep{Liubarskii83, Chluba2008c}. It is thus important to understand how the CRR depends on different assumptions, and a detailed computation of the signals for the standard scenario is the first step.

Several calculations of the CRR have been carried out in the past using different methods and approximations \citep[e.g.,][]{Dubrovich1975, Rybicki93, DubroVlad95, Dubrovich1997, Burgin2003, Kholu2005, Wong2006, Jose2006, Chluba2007}. In Sect.~\ref{sec:method}, we dwell deeper into the details, but the biggest problem is that these calculations either included a small number of atomic levels, neglected physical processes and/or took prohibitively long to allow detailed computations for many cosmologies. The situation was very similar back in the early days of CMB cosmology when slow Boltzmann codes \citep{Ma1995, Hu1995} were used for computations of the CMB power spectra.

In this work, we take the next step for computations of the CRR, developing the software package {\tt CosmoSpec}\footnote{{\tt CosmoSpec} is based on {\tt CosmoRec} \citep{Chluba2010b}, with additional modules to allow computation of the recombination radiation. It will be made available at \url{www.Chluba.de/CosmoSpec}}, which allows fast and precise computation of the CRR for a wide range of cosmological models including all important physical processes. The computational method is based on a combination of the effective multilevel atom\footnote{This method was independently proposed by \citet{Burgin2003}.} \citep{Yacine2010} and effective conductance approach \citep{Yacine2013RecSpec}, which we briefly review in Sect.~\ref{sec:method}. These approaches are similar in spirit to the line-of-sight approach for CMB anisotropies \citep{CMBFAST} used in modern Boltzmann codes {\tt CMBfast}, {\tt CAMB} \citep{CAMB} or {\tt CLASS} \citep{CLASSCODE}, in the sense that they rely on a factorization of the problem into a slow but cosmology-independent part and a fast, cosmology-dependent part.

With {\tt CosmoSpec} the calculation of the CRR for one cosmology can be performed in $\simeq 15$ seconds on a standard laptop, a performance that opens the path for detailed parameter explorations and experimental forecasts, applications we leave to future work.
We discuss the importance of different physical processes, such as interspecies feedback, collisions and uncertainties in the atomic model for neutral helium (Sect.~\ref{sec:distortions}). Overall, {\tt CosmoSpec} in its current state should be accurate at the level of a few percent for the standard recombination scenario, mainly due to uncertainties of the neutral helium atom model. 

\vspace{-5mm}
\section{Atomic models}
\label{sec:method}
For the computations presented here, we adopt atomic models for hydrogen and helium closely following \citet{Jose2006, Chluba2007} and \citet{Jose2008}. Below we briefly summaries the main aspects of these models, but refer to the above references for details.

\vspace{-2mm}
\subsection{Hydrogen and hydrogenic helium}
Level energies, electric dipole transitions rates, and photoionization cross sections for hydrogenic ions can be computed analytically using the expressions of \citet{Karzas1961} and the recursion method of \citet{StoreyHum1991} \citep[see also][]{Hey2006}. Corrections due to the Lamb shift are neglected. We also omit \ion{H}{i} and \ion{He}{ii} quadrupole transitions, which coincide with the dipole transition frequencies and have been shown to cause a negligible effect on the recombination dynamics \citep{Grin2009, Chluba2012HeRec} and thus should not change the \ion{H}{i} and \ion{He}{ii} recombination radiation significantly. 

We use the hydrogen radiative transfer module of {\tt CosmoRec} \citep{Chluba2010b}, resolving the 2s, 2p, 3s, 3p and 3d states. This includes, Lyman-$\alpha$ and $\beta$ resonance scattering as well as two-photon emission and Raman scattering events. Electron scattering is added using a Fokker-Planck approach \citep{Hirata2009, Chluba2009b} to improve the stability of the transfer calculation, but we neglect modifications caused by the detailed shape of the scattering kernel \citep{Yacine2010b}. The results of the transfer calculation are used to obtain the high-frequency distortion caused by hydrogen. The \ion{H}{i} 2s-1s two-photon decay rate is set to $A^{\rm HI}_{\rm 2s1s}=8.2206\,\sec^{-1}$ \citep{Labzowsky2005}.

For hydrogenic helium, we approximate the 2s-1s two-photon profile using the expression of \citet{Nussbaumer1984}:
\beal
\label{eq:phi_appr}
\phi(y)=C\left[w(1-4^\gamma\,w^\gamma)+\alpha\,w^{\beta+\gamma}\,4^\gamma\right]
\Abst{,}
\end{align}
where $y=\nu/\nu_{\rm 2s1s}$, $w=y[1-y]$, $C=12.278$, $\alpha=0.88$, $\beta=1.53$ and $\gamma=0.8$. The two-photon profile is normalized as $\int \phi(y) \id y = 2$, since two photons are emitted per transition. We set the \ion{He}{ii} two-photon decay rate to\footnote{We obtained this value by applying the scaling $A^Z_{\rm 2s1s}\propto Z^6$ for charge $Z=2$ with relativistic corrections from \citet{Drake1986} and renormalizing to $A^{\rm HI}_{\rm 2s1s}=8.2206\,\sec^{-1}$ for \ion{H}{i} (i.e., $Z=1$). In previous computations, $A^{\rm HeII}_{\rm 2s1s}=526.5\,\sec^{-1}$ was used, but the difference is small.} $A^{\rm HeII}_{\rm 2s1s}=526.57\,\sec^{-1}$. For $\ion{He}{ii}$, we use an effective multilevel atom approach with Sobolev escape probability formalism. No separate radiative transfer calculation is performed.

\vspace{-3mm}
\subsection{Neutral helium}
The atomic model for helium significantly more complex than the one of hydrogenic ions and, more importantly, more uncertain. The main complication is due to the fact that the neutral helium atom is a three particle system (nucleus + two electrons), so that no closed analytic solutions exist. 

\vspace{-3mm}
\subsubsection{Energies and transitions}
For the energies and transitions rates among the singlet and triplet levels with principal quantum numbers $n\leq 10$ we use the multiplet tables from \citet{Drake2007}. These tables have a few gaps (e.g., for level $\{n, l, s, j\}=\{10, 7, 0, 7\}$), which we fill with hydrogenic values (see Appendix~\ref{app:transitions} for details). To obtain the energies of levels with $10<n\leq 30$ and $l\leq6$ we use the quantum defect method \citep[e.g., see][]{DrakeBook2006}, while for all other levels we use the simple hydrogenic approximation for the level energy, $E_n\approx -E_{\rm He}/n^2$, where $E_{\rm He}=hc\,R_\infty/(1+\me/M_{\rm He})$, $\me/M_{\rm He}\approx \pot{1.3709}{-4}$ is the electron to helium mass ratio\footnote{We simply use the mass of an $\alpha$-particle for $M_{\rm He}$.} and $R_\infty\approx \pot{1.0974}{7}\,{\rm m^{-1}}$ is the Rydberg constant. For levels with $n>10$, we generally do not include $J$-resolved levels. For transitions between levels with $n>10$, we use hydrogenic transitions rates, rescaled by the second power\footnote{The frequency rescaling is only significant for the $n^1 {\rm P} - 1^1{\rm S}$ series.} of the transitions frequency. This approximation is rather crude for transitions involving low-$\ell$ states (i.e., $n$S through $n$D) as illustrated below (Sect.~\ref{sec:uncertain_He}).
No singlet-triplet transitions are allowed among excited states with $n>10$. For transitions from excited levels to $n'\leq10$, we use hydrogenic values following \citet{Jose2008}, which have the correct average property (Appendix~\ref{app:transitions}). We also include a few extra transitions  \citep[see][for more details]{Chluba2009c, Chluba2012HeRec}, for example, singlet-triplet transitions \citep{Switzer2007I} and additional \ion{He}{i} quadrupole lines \citep{Cann2002} to the ground-state. The radiative transfer problem for \ion{He}{i} is solved using {\tt CosmoRec} \citep[for details see,][]{Chluba2012HeRec}.

We approximate the \ion{He}{i} $\HeIlevel{2}{1}{S}{0}-\HeIlevel{1}{1}{S}{0}$ two-photon profile using \citep{Drake1986, Switzer2007I}:
\beal
\label{eq:phi_appr}
\phi(y)=C\,\frac{w^3}{(w+\delta)^2}\,\left[\alpha-\beta w+\gamma w^2\right]
\Abst{,}
\end{align}
where $y=\nu/\nu_{\HeIlevel{2}{1}{S}{0}-\HeIlevel{1}{1}{S}{0}}$, $w=y[1-y]$, $C=19.604$, $\alpha=1.742$, $\beta=7.2$, $\gamma=12.8$ and $\delta=0.03$. Again, the two-photon profile is normalized as $\int \phi(y) \id y = 2$. We use $A^{\rm HeI}_{\rm 2s1s}=51.3\,\sec^{-1}$ \citep{Drake1969} for the total two-photon decay rate. The effect of other two-photon and Raman processes \citep{Switzer2007III} on the recombination dynamics and shape of the high-frequency distortion are neglected. We also neglect corrections caused by the detailed photon redistribution kernel of electron and line scattering \citep{Switzer2007II, Chluba2012HeRec}.

\vspace{-3mm}
\subsubsection{Photoionization cross section}
As pointed out previously \citep[e.g.,][]{Jose2008, Chluba2009c, Glover2014}, for the CRR one of the biggest uncertainty in the neutral helium model is due to the lack of data for the photoionization cross sections. For levels with $n\leq 10$ and $l\leq 3$, we use a combination of the sparse {\tt TopBase} data \citep{Cunto1993} and pre-calculated recombination rates from \citet{Smits1996}. For all other levels, we use hydrogenic approximations for the photoionization cross section, following \citet{Jose2008}. The $J$-resolved photoionization cross section is simply set to $\sigma^{\rm He}_{nLJ}(\nu)\approx \sigma^{\rm H}_{nL}(\nu_{i\rm c}^{\rm H}\, \nu/\nu_{i\rm c}^{\rm He})$ using the ionization threshold frequencies for hydrogen and helium, $\nu_{i\rm c}^{\rm H}$ and $\nu_{i\rm c}^{\rm He}$. For the recombination rate this implies a rescaling by $\simeq (\nu_{i\rm c}^{\rm He}/\nu_{i\rm c}^{\rm H})^3$. This approximation is obtained by assuming that the $J$-averaged cross section is equal to the hydrogenic one and using the fact that $J$-resolved cross sections are all equal, up to small corrections due to the fine splitting of $J$-sub-levels (see Appendix \ref{app:sigma}).\footnote{Another approximation, $\sigma^{\rm He}_{nLJ}\approx (2J+1)\sigma^{\rm H}_{nL}/[(2L+1)(2S+1)]$, is used by \citet{Bauman2005}, however, this does not give the correct $J$-averaged value, $\sigma^{\rm He}_{nL}=\sum_J (2J+1)\sigma^{\rm He}_{nLJ}/[(2L+1)(2S+1)]\neq \sigma^{\rm H}_{nL}$. With this approximation, we find that the low-frequency helium recombination emission increases by $30-40\%$, emphasizing how important knowledge of the detailed recombination rate is for the computation of the recombination spectrum.} 
It also neglects corrections to the shape of the cross section, in particular due to auto-ionization resonances, which are important for the low-$\ell$ states. This leads to uncertainties in the shape of the free-bound spectrum, as we briefly discuss in Sect.~\ref{sec:uncertain_He}.

\vspace{-3mm}
\section{Computational method}
\label{sec:method}

\subsection{Effective multilevel atom and conductance methods}
One of the biggest challenges for detailed and fast computation of the cosmological recombination history and radiation is that in explicit multilevel approaches \citep[e.g.,][]{Seager2000, Chluba2007} the populations of many levels ($\simeq n_{\rm max}^2/2$ for hydrogen) have to be followed to determine their departures from full equilibrium. For 500-shell calculations of hydrogen this includes some $\simeq 10^5$ levels, which makes the computations very time-consuming \citep[e.g.,][]{Jose2006, Grin2009, Chluba2010}. However, the effect of excited states can be integrated out using the effective multilevel atom method for the recombination history  \citep{Yacine2010} and the effective conductance method for the recombination radiation \citep{Yacine2013RecSpec}.

The effective multilevel atom\footnote{The method is perhaps more accurately described as an effective \emph{few-level} atom.} is an \emph{exact} generalization of the effective three-level atom model \citep{Peebles68, Zeldovich68}. In this approach, the case-B recombination coefficient is replaced by effective recombination coefficients $\mathcal{A}_i(\Te, \Tg)$ to a few low-lying excited states. These coefficients account for stimulated recombinations and the finite probability of photoionization as a captured electron cascades down to the lowest excited states. In the absence of collisional transitions, they are functions of the electron ($\Te$) and photon ($\Tg$) temperatures only. These coefficients are supplemented by effective photoionization rates $\mathcal{B}_i(\Tg)$ and effective transition rates between the low-lying excited states, $\mathcal{R}_{i\rightarrow j}(\Tg)$. The calculation of the effective rates to high accuracy is computationally demanding, as it requires solving large linear algebra problems. However, it has to be done only once. For any cosmology, the recombination history can then be computed very efficiently by solving an effective few-level system, and interpolating the effective rates whenever they are needed. The effective multilevel approach is the basis for the cosmological recombination codes {\tt HyRec} \citep{Yacine2010c} and {\tt CosmoRec} \citep{Chluba2010b} which are now directly used for the analysis of high-precision CMB data from {\it Planck} \citep{Planck2015params}.

The effective conductance method \citep{Yacine2013RecSpec} allows to factorize the computation of the recombination radiation into a slow, cosmology-independent part, and a fast, cosmology-dependent part. In this approach, the net emissivity (or intensity) in any bound-bound or free-bound transition is rewritten as a sum of temperature-dependent coefficients (effective conductances) times departures from equilibrium of the populations of the lowest-lying excited states (which play the role of tensions). The former are pre-tabulated as a function of transition energy and temperature, while the latter are obtained very efficiently, for any particular cosmology, by solving the effective few-level atom described above.

Here, we extend {\tt CosmoRec} to allow detailed computation of the $\ion{He}{iii}\rightarrow\ion{He}{ii}$ recombination history using an effective multilevel approach. Similarly, we update the effective rate coefficients for neutral helium to include 500 shells. While the effective rates and conductances for neutral helium have to be explicitly computed, those for hydrogenic helium can be obtained from those for hydrogen by simple rescaling which we derive in Appendix \ref{app:rescalings}.

\vspace{-3mm}
\subsection{Electron scattering}
Line broadening and shifting caused by scattering with free electrons was found to be important for the \ion{He}{ii} recombination radiation \citep{Jose2008}. To include these effects we use the improved approximation for the electron scattering Green's function of the photon occupation number given by \citet{Chluba2015GreensII}:
\beal
\label{eq:Dn_sol_ZS_improved_II}
G(x', x, y)&=
\frac{\exp\left(-\left[\ln(x/x')-\alpha \,y+\ln(1+x' y)\right]^2/4y \,\beta \right)}{\sqrt{4\pi y \,\beta}\,x^3}.
\end{align}
Here, $x=h\nu/k\Tg$ and $y=\int (k\Te/\me c^2)\, \sigT \Ne c \id t$ denotes the scattering $y$-parameter. We also introduced $\alpha=[3-2 f(x')]/\sqrt{1+x' y}$ and $\beta=(1+x' y[1-f(x')])^{-1}$, with $f(x)=\expf{-x}(1+x^2/2)$, which were obtained by matching the numerical solution. This approximation improves over the classical solution ($\alpha=3$, $\beta=1$ and $\ln(1+x' y)\rightarrow 0$) of \citet{Zeldovich1969} by accounting for the effect of electron recoil at high frequencies and including the blackbody-induced stimulated scattering \citep{Chluba2008d} at low frequencies.

To compute the contribution to frequency bin $x_j$ caused by the scattering of photons emitted at frequency $x_i$ and redshift $z$, we find
\beal
\label{eq:DI_j_sc}
\Delta I^{\rm sc}_j&=\frac{\Delta I_i(z)}{2\,x_i}\Big[{\rm erf}(b_u)-{\rm erf}(b_l)\Big]\expf{-a_i+y \,\beta_i}
\nonumber\\
&\approx \frac{\Delta I_i(z)}{x_i}\,\frac{\expf{-a_i+y \,\beta_i-b_j^2}}{\sqrt{4\pi y\, \beta_i}}\,\frac{x_u-x_l}{x_j}.
\end{align}
We integrated across the bin between $x_l$ and $x_u$ with $x_j=(x_u+x_l)/2$. Here, $a_i=-\ln x_i-\alpha(x_i) \,y+\ln(1+x_i y)$, $b_k=(a_i-2y\, \beta_i+\ln x_k)/2\sqrt{y\, \beta_i}$ and $\Delta I_i(z)$ denotes the distortion caused by the process without including electron scattering. The second line in Eq.~\eqref{eq:DI_j_sc} was obtained assuming a narrow bin $x_u-x_l\ll x_j$. The $y$-parameter is computed between the emission redshift $z$ and today using the solution for the electron recombination history of {\tt CosmoRec}. Around the maximum of emission from \ion{He}{ii} ($z\simeq6000$) we find $y\simeq 10^{-3}$, so that the line broadening reaches $\Delta \nu/\nu\simeq 2 \sqrt{y \ln2}\simeq 5\%$. During neutral helium and hydrogen recombination, electron scattering is much less important and can safely be neglected.

\vspace{-0mm}
\subsection{Free-free absorption}
At low frequencies ($\nu\lesssim 0.1-1\,\GHz$), free-free absorption becomes noticeable \citep{Chluba2007}. So far, this effect was only taken into account for the \ion{H}{i} recombination radiation. It is, however, slightly more important for the helium recombination spectra, as we show below (Sect.~\ref{sec:distortions}). To include this process, we modify the emission in each frequency bin $x_i$ and redshift $z$ by
\beal
\label{eq:DI_i_ff}
\Delta I^{\rm ff}_i&=\Delta I_i(z)\,\expf{-\tau_{\rm ff}(x_i, z)},
\end{align}
where the free-free absorption optical depth is roughly given by, $\tau_{\rm ff}(x, z)\approx F(z) \ln(2.25/x)/x^{2}$, and $F(z)$ is a single redshift-dependent function. This function can be computed using approximations for the free-free Gaunt factors \citep[e.g.,][]{Itoh2000, Draine2011Book}. The efficiency of free-free absorption drops exponentially around hydrogen recombination ($z\simeq 10^3$), but earlier it is important at low frequencies. More details can be found in \citet{Chluba2015GreensII}.

\vspace{-0mm}
\subsection{Feedback due to helium photons}
\label{sec:HeII_feedback_method}
The feedback physics is simple: energetic photons, emitted during the two recombination phases of helium, can {\it ionize} and {\it excite} the next-lying species at a later time. Specifically, $\ion{He}{ii}$ photons feedback onto \ion{He}{i} and \ion{H}{i} atoms, and \ion{He}{i} photons feedback onto \ion{H}{i} atoms. These ionizations and excitations typically take place earlier than the epoch of recombination of the lower-lying species, and are instantaneously compensated by rapid recombinations and decays, generating a pre-recombination radiation from the species that is subjected to feedback. In the case of $\ion{He}{ii}\rightarrow \ion{He}{i}$ feedback, the pre-recombination emission from $\ion{He}{i}$ can itself feedback onto hydrogen \citep{Chluba2009c, Chluba2012HeRec}.

Part of the $\ion{He}{i}\rightarrow \ion{H}{i}$ feedback is already included in previous calculation for the hydrogen recombination radiation \citep{Jose2008}. However, as shown in \citet{Chluba2009c}, changes in the recombination dynamics due to \ion{He}{ii} feedback cause significant additional modifications of the CRR. 
In \citet{Chluba2009c}, these corrections were only included for 20-shell calculations, which here we extend to 500 shells. Furthermore, changes to the detailed time-dependence of the feedback process because of radiative transfer effects (e.g., broadening by electrons scattering of the $\ion{He}{ii}$ Lyman-$\alpha$ line) were neglected. These effects can be readily included in our new computation. For this we modify the radiative transfer calculation of neutral helium, including the emission of photons in the $\ion{He}{ii}$ Lyman-$\alpha$ line and 2s-1s two-photon continuum. These photons then redshift and feedback on neutral helium and hydrogen in their respective pre-recombination eras causing the production of extra photons. Our treatment automatically includes corrections due to line broadening by electron and resonance scattering. The results of these calculations will be presented in Sect.~\ref{sec:detailed_feedback}.

\vspace{-0mm}
\subsection{Collisional transitions}
\label{sec:colls}
Collisional processes can modify the atomic level populations of the highly excited states. This affects the recombination dynamics and the low-frequency emission, originating from transitions involving highly excited states ($n\gtrsim 30-50$). For hydrogen, collisional processes were found to have a minor effect for both aspects\footnote{For the recombination radiation from hydrogen, collisions become noticeable at $\nu\lesssim0.1\,\GHz$ \citep[Fig.~11 of][]{Chluba2010}, but here we shall restrict ourselves to $\nu\gtrsim 0.1\,\GHz$.} \citep{Chluba2007, Chluba2010}, however, because collisions become more important at higher densities, for helium this may no longer be true, in particular for the pre-recombination emission. 

\vspace{-0mm}
\subsubsection{Angular-momentum-changing collisions}
Angular-momentum-changing collisions are the most important for the recombination history and the CRR. To estimate the effect of $\ell$-changing collision among energetically degenerate levels (modeling them with hydrogenic wave functions), we use the classical approximation of \citet{Vrinceanu2012}, which we extend to general ion and nucleus charge:
\beal
\label{eq:C_ul_ell}
C_{n\ell \rightarrow n \ell'}&\approx\pot{2.6}{-5}\sqrt{\frac{M}{\me}}\,\frac{(Z^2_{\rm ion}/Z^2_{\rm nuc})\,N_{\rm ion}}{\sqrt{\Te/\Kel}}
\nonumber\\
&\qquad\times \frac{n^2\left[n^2(\ell+\ell')-\ell^2_{<}(\ell+\ell'+2|\Delta \ell|)\right]}{(2\ell +1) \,|\Delta\ell|^3}
\,\cm^3\,{\rm s}^{-1},
\end{align}
where $N_{\rm ion}$ is the number density of the projectiles, $\ell_{<} = \min(\ell, \ell')$, $M$ is the reduced mass of the ion-target system; $Z_{\rm ion}$ the charge of the projectile and $Z_{\rm nuc}$ that of the nucleus ($Z_{\rm nuc} = 1$ for hydrogen and neutral helium, and 2 for singly-ionized helium). We give a simple derivation of Eq.~\eqref{eq:C_ul_ell} in Appendix \ref{app:coll}, valid for $\ell, \ell', n \gg 1$. We also argue that the underlying assumption of a straight-line trajectory for the colliding ion is very accurate at the relevant temperatures, even when the target is charged as is the case for singly-ionized helium. For $\ell$-mixing, proton and $\alpha$-particle collisions are more important than electron collisions (which we neglect). 

Equation~\eqref{eq:C_ul_ell} automatically obeys detailed balance and provides an estimate for transition with\footnote{In practice, we include up to $|\Delta \ell|=3$, which is sufficient.} $|\Delta \ell|>1$, in very good agreement with the full quantum calculation \citep{Vrinceanu2012}. It can be applied for all three species, since $\ell$-changing collisions are only important for highly-excited states ($n \gtrsim 30-50$), which even for neutral helium are close to hydrogenic. It does, however, omit the possibility of collision-induced singlet-triplet state mixing. 

For $\Delta \ell = \pm 1$, the quantum-mechanical transition probability scales as $1/b^2$ for large impact parameters, while the classical transition probability used to obtain Eq.~\eqref{eq:C_ul_ell} vanishes beyond a maximum impact parameter \citep[see Fig.~1 in][as well as our Appendix \ref{app:coll}]{Vrinceanu2012}. To obtain a finite collision rate with the full quantum description, one therefore needs to cut off the integration at some maximum impact parameter \citep{Vrinceanu2001}. Since the divergence is only logarithmic, the choice of the cutoff is not critical and the result is found to be well approximated by Eq.~\eqref{eq:C_ul_ell} (Sadeghpour, 2015, private communication). It is lower by a factor of a few than the previous estimate of \citet{Pengelly1964}, whose Born approximation calculation required a regularization at both large and small impact parameters. 

\vspace{-0mm}
\subsubsection{Collisional excitations and de-excitations}
For collisional de-excitation rates of ions or atoms by electron impact we use the expression of \citet{Rege1962}:
\beal
\label{eq:C_ul}
C_{i\rightarrow j}\approx20.6\,\Ne \,\lambda_{ij}^3\,\frac{A_{i\rightarrow j}}{\sqrt{\Te/\Kel}}\,P(h\nu_{ij}/k\Te),
\end{align}
where $i$ and $j$ are the the upper and lower levels, respectively, $\lambda_{ij}=c/\nu_{ij}$ is the transition wavelength, $A_{i \rightarrow j}$ is the Einstein-A coefficient of the transition and the function $P(y)$ depends on whether the target is neutral or positive. At high temperatures, it scales as $P(y)\approx -0.159 - 0.276 \ln y$ in both cases. The collisional excitation rate is obtained by detailed balance: $C_{j\rightarrow i} = (g_i/g_j) \expf{-h\nu_{ij}/k\Te} C_{i \rightarrow j}$.

\vspace{-0mm}
\subsubsection{Collisional ionizations and three-body recombinations}
For the collisional ionization rate by electron impact, we use the simple expression \citep[see,][]{Mihalas1978, Mashonkina1996}:
\beal
\label{eq:C_ic}
C_{i\rightarrow \rm c}\approx\pot{1.55}{13}\,\bar{g}\,\Ne \,\frac{\sigma_{i \rightarrow \rm c}(\nu_{i \rm c})}{\sqrt{\Te/\Kel}}\,\frac{\expf{-h\nu_{i \rm c}/k\Te}}{h\nu_{i \rm c}/k\Te} \, \cm\,{\rm s}^{-1},
\end{align}
where $\sigma_{i \rightarrow \rm c}(\nu_{ic})$ is the photoionization cross section at threshold, $\bar{g}=0.2$ for hydrogen and neutral helium and $\bar{g}=0.3$ for hydrogenic helium. The collisional recombination coefficient follows from detailed balance. 

The rate coefficients for collisional ionization and excitation are uncertain. However, they turn out to have a very small effect on the CRR, as we shall discuss in Sect.~\ref{sec:colls_discussion}.

\subsubsection{Inclusion in the effective rates and conductances}

One can simply generalize the effective multilevel and effective conductance methods to account for collisional transitions. The effective rates and conductances are then not only functions of temperature, but also the collider densities, $N_{\rm coll} = \{ N_{\rm e}, N_{\rm p}, N_{\rm He}^+, N_{\rm He}^{++}\}$. In principle this adds more dimensions over which to tabulate the effective rates. However, since the cosmological parameters are known to percent accuracy, so is the evolution of the main collider densities as a function of $\Tg$. It would therefore probably be enough to compute the values and first few derivatives of the effective rates with respect to the densities around their best-fit values $N_{\rm coll}^0(\Tg)$. However, given that the uncertainty of the collisional rates is still large compared to the precision of the standard cosmological parameters, in this paper we simply compute the effective rates and conductances at the best-fit collider densities $N_{\rm coll}^0(\Tg)$.

\section{Results}
\label{sec:distortions}
After obtaining all the effective rates and conductances, both the computation of the recombination history and recombination radiation boils down to solving a moderate system of coupled ordinary differential and partial differential equations and numerical integrals. This accelerates the calculation by a large amount, making it feasible to perform one full calculation within $\simeq 15$ seconds on a standard laptop. Here, we illustrate the results for the recombination radiation, highlighting some of the new aspects and comparison with previous computations. 

\subsection{Comparison with previous computations}
To validate our new treatment, we first compare our results with previous calculations of the recombination radiation. For hydrogen, detailed multilevel recombination computations for the bound-bound emission, including up to 350 $\ell$-resolved shells ($\simeq 61,000$ individual levels), were presented in \citet{Chluba2010}. These calculations also included the effect of $\ion{H}{i}$ Lyman-continuum absorption during the $\ion{He}{i}$ recombination era \citep{Kholupenko2007, Switzer2007I}, an effect that leads to pre-recombination emission from hydrogen \citep{Jose2008, Chluba2009c}. In \citet{Yacine2013RecSpec}, the effective conductance method was used to compute the recombination radiation from hydrogen for 500 $\ell$-resolved shells, also including the free-bound radiation, which previously was only available for $100$-shell hydrogen calculations \citep{Chluba2006b}. Comparing with the results from our new computation for similar settings (in terms of the included physics and atomic model), we find excellent agreement at the level of $\simeq 0.1\%-1\%$ with these previous computations. While this validates our implementation, here we add  corrections due to feedback from $\ion{He}{ii}$ photons, which increases the emission caused by hydrogen and neutral helium (see Sect.~\ref{sec:detailed_feedback}).

For singly-ionized helium, we compare with the multilevel computations presented in \citet{Jose2008} for 100 shells. Only results for the bound-bound radiation were previously presented, for which we find excellent agreement both with and without electron scattering included. For the neutral helium recombination emission only the bound-bound emission for up to 30 shells was previously considered \citep{Jose2008}. Extending the multilevel code of \citet{Chluba2010} to include more levels for neutral helium, we confirmed the results of our conductance method. For the bound-bound radiation,  at low frequencies ($\nu\lesssim 20\,\GHz$) our results deviate from the spectrum presented in \citet{Jose2008}. We attribute these differences to variations in the atomic model of neutral helium (in particular the recombination rates to excited states), however, the main high-frequency features agree well with our improved calculations. Here we also add the two free-bound contributions from both helium ions, which have not been obtained before.

\begin{figure}
\centering 
\includegraphics[width=\columnwidth]{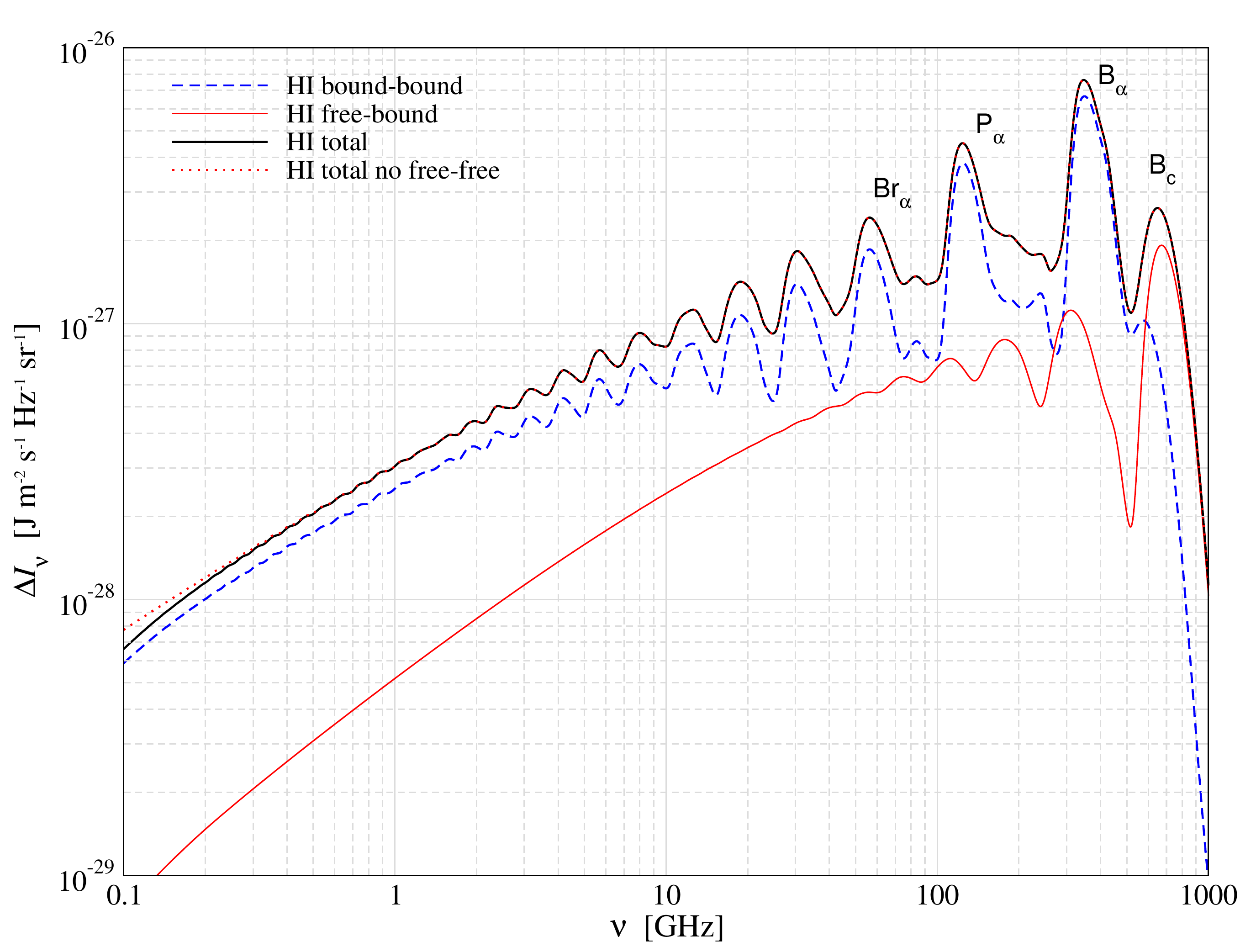}
\\[-1mm]
\includegraphics[width=\columnwidth]{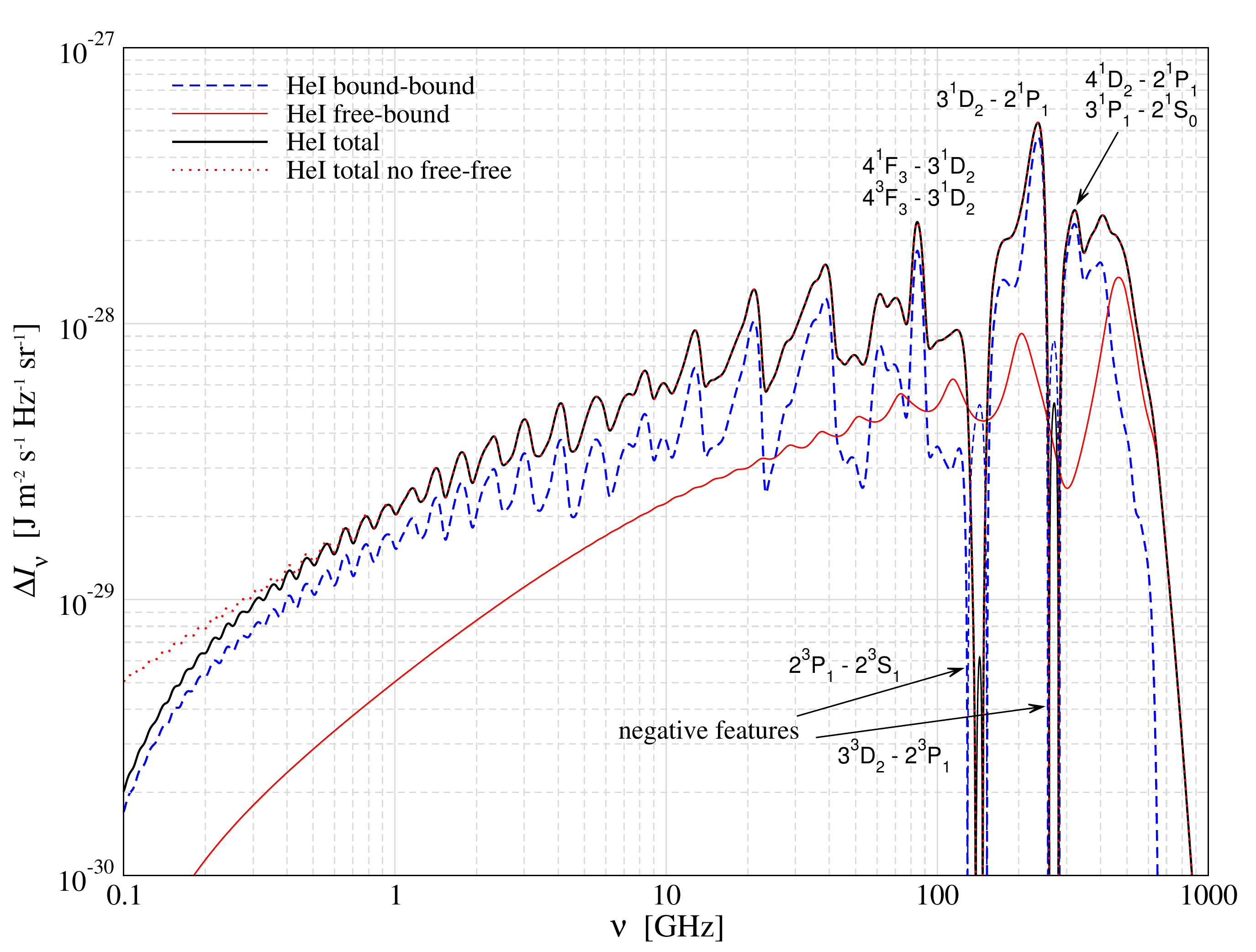}
\\[-1mm]
\includegraphics[width=\columnwidth]{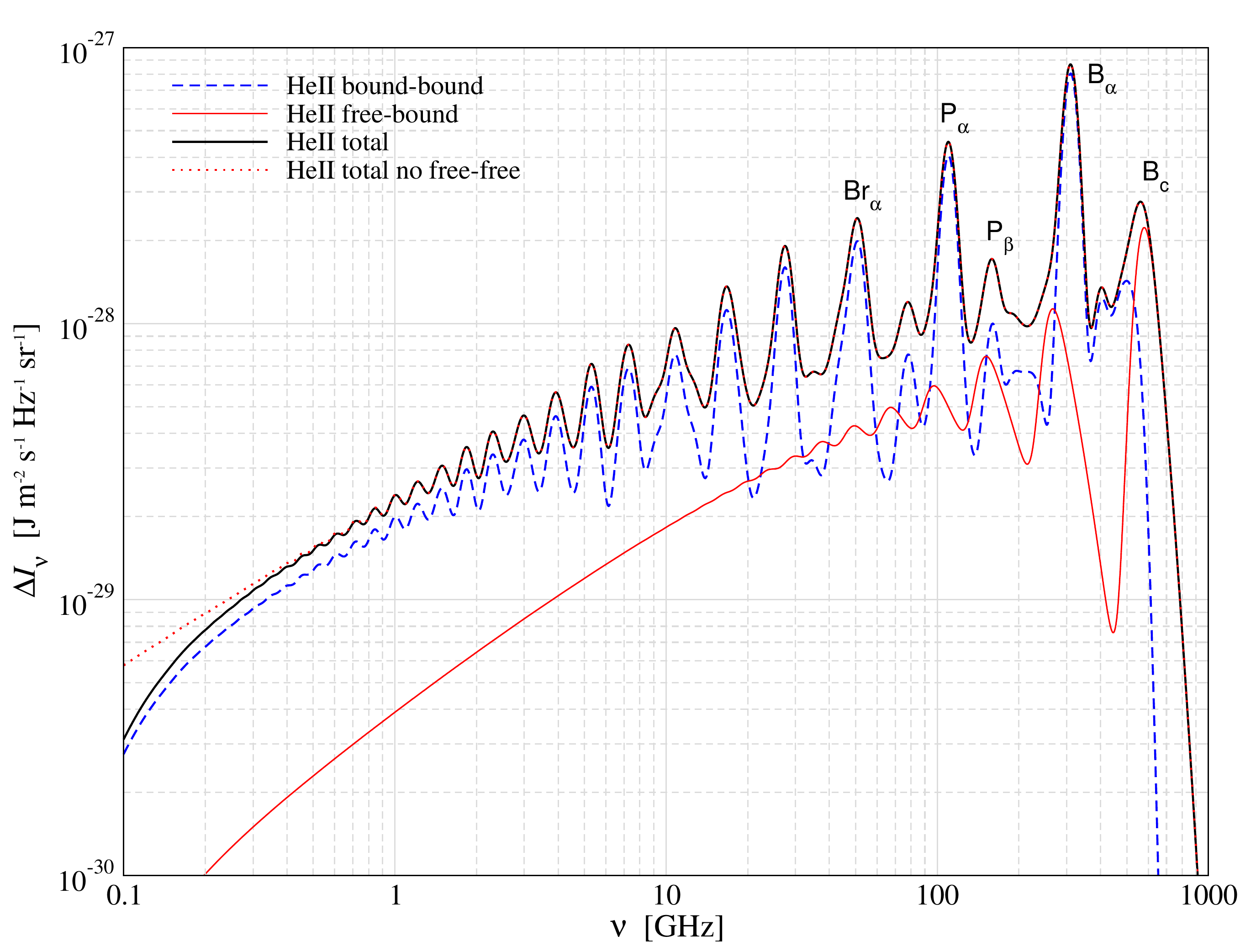}
\caption{Cosmological recombination radiation from hydrogen and helium for 500 shells. $\ion{He}{ii}$ feedback and collisions were neglected and only the emission among levels with $n\geq 2$ is shown. For the total spectra we also illustrate the effect of free-free absorption.}
\label{fig:He-spectra}
\end{figure}

\vspace{-2mm}
\subsection{Contributions from the three recombination eras}
\label{sec:detailed_feedback}
In Fig.~\ref{fig:He-spectra}, we illustrate the individual contributions from bound-bound and free-bound transitions of the three atomic species. We only show the emission among levels with $n\geq 2$, and do not display the high-frequency distortion from the Lyman lines and and 2s-1s continuum emission, though we compute them consistently as as part of the radiative transfer problem. The effect of collisions and \ion{He}{ii} feedback were neglected to produce Fig.~\ref{fig:He-spectra} and will be discussed below.

The hydrogen recombination radiation shows several feature (Balmer, Paschen and Brackett lines) related to the bound-bound $\alpha$ transitions ($\Delta n=1$), with the emission merging to a quasi-continuum at low frequencies. Free-free absorption is noticeable at $\nu\lesssim 0.3\,\GHz$. Although on average at a $\simeq 13$ times lower level, the distortions from the two helium eras also show significant structure. For the \ion{He}{ii} spectrum, the effect of electron scattering is important, but nevertheless distinct features remain visible. The total \ion{He}{i} spectrum, shows absorption features at $\nu \simeq 145\,\GHz$ and $\nu\simeq 270\,\GHz$. The first feature is related to the 10\,830 $\AA$ line ($
\leftrightarrow 2^3{\rm P}_1-2^3{\rm S}_1$) and the second to absorption in the 3D-2P triplet-triplet transitions \citep[see][for more details]{Jose2008}. For the helium contributions, free-free absorption becomes significant at $\nu\lesssim 0.5\,\GHz$, exponentially cutting off all recombination emission at $\nu\lesssim 50-100\,{\rm MHz}$.

\begin{figure*}
\centering 
\includegraphics[width=2\columnwidth]{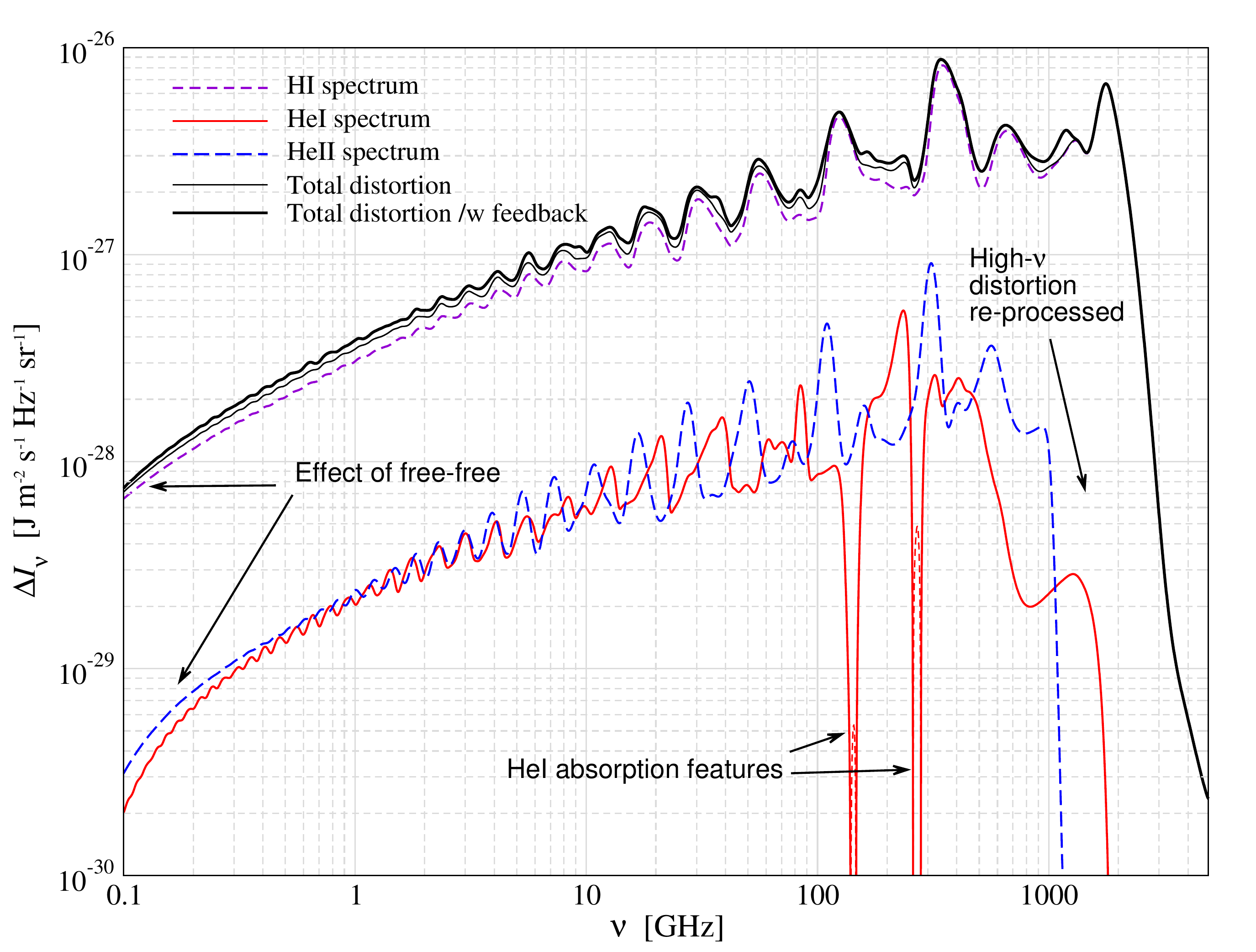}
\caption{Cosmological recombination radiation from hydrogen and helium for 500-shell calculations. The different curves show individual contributions (without feedback) as well as the total distortion with and without feedback processes. At low frequencies, free-free absorption becomes noticeable. The effect is stronger for the contributions from helium due to the larger free-free optical depth before recombination ends at $z\simeq 10^3$. In total, some $6.1\,\gamma$ are emitted per hydrogen atom when all emission and feedback are included. Hydrogen alone contributes about $5.4\,\gamma/N_{\rm H}$ and helium $0.7\,\gamma/N_{\rm H}$ ($\simeq 8.9\,\gamma/N_{\rm He}$).}
\label{fig:total}
\end{figure*}

In Figure~\ref{fig:total}, we show the total (bound-bound and free-bound) recombination radiation from hydrogen and helium for our detailed 500-shell calculations. We also included the high-frequency distortion, which for the helium spectra is strongly absorbed and reprocessed, reappearing as pre-recombination hydrogen Lyman-$\alpha$ emission \citep{Jose2008, Chluba2012HeRec}. Overall, the effect of helium is visible in several bands, introducing features, shifts in the line positions and enhanced emission. In particular, when adding all feedback corrections (see below for more discussion), the effect of helium becomes significant, in some parts well in excess of the expected average $\simeq 16\%$ contribution. This will help significantly when attempting to use the CRR to determine the pre-stellar (un-reprocessed) helium abundance.

\vspace{-3mm}
\subsection{Effect of \ion{He}{ii} feedback}
\label{sec:HeII_feedback_phys}
\begin{figure}
\centering 
\includegraphics[width=\columnwidth]{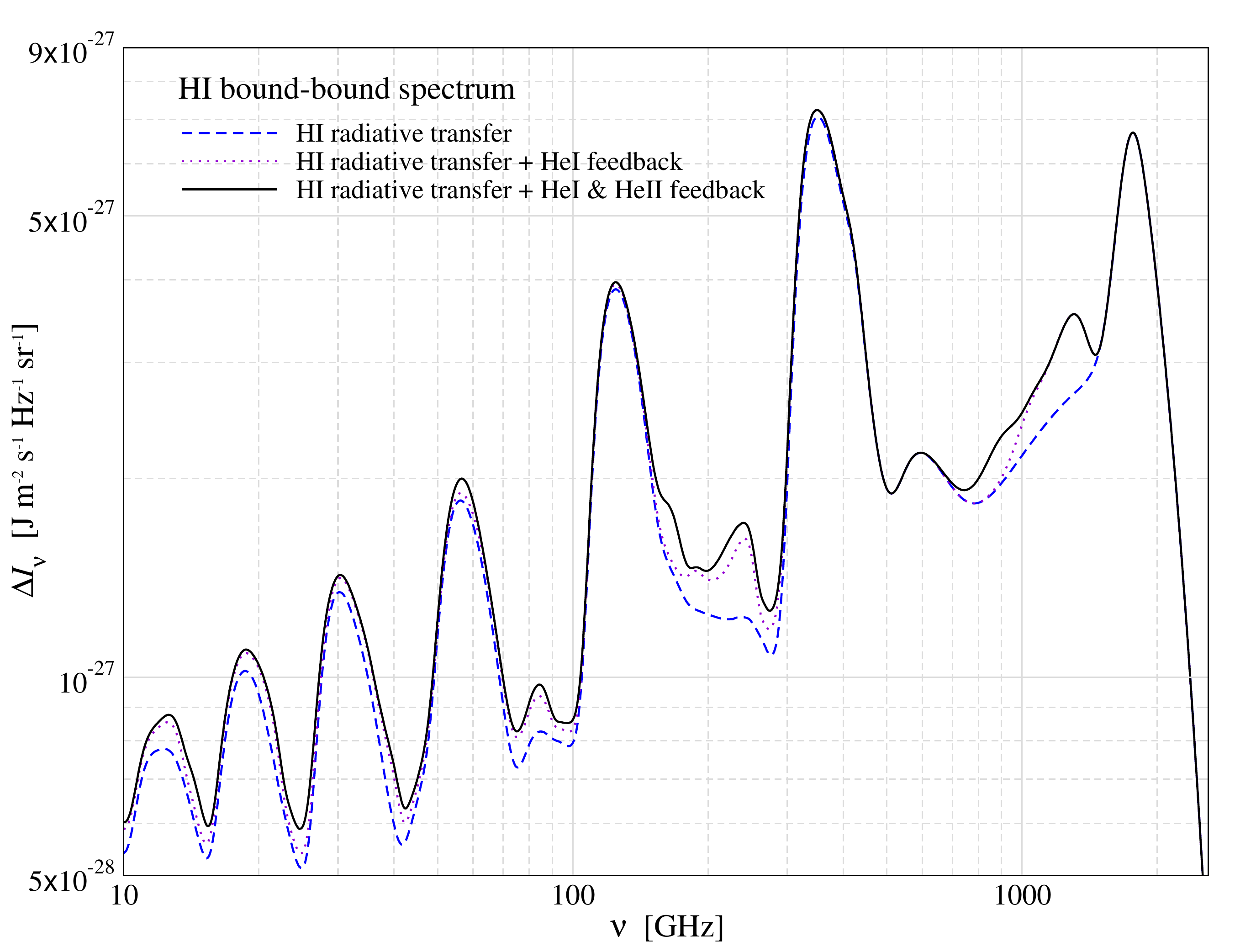}
\caption{Helium feedback corrections to the \ion{H}{i} bound-bound recombination radiation. The conductances were computed for $\nmax=500$. The main features due to helium feedback are in good agreement with the 20-shell calculations of \citet{Chluba2009c}.}
\label{fig:HI_BB_feedback}
\end{figure}

As explained in Sect.~\ref{sec:HeII_feedback_method}, feedback caused by high-frequency photons emitted during the $\ion{He}{iii} \rightarrow\ion{He}{ii}$ recombination era changes the CRR of hydrogen and neutral helium by ionizing neutral atoms in their pre-recombination eras \citep{Chluba2009c}. In Fig.~\ref{fig:HI_BB_feedback}, we illustrate the effect on the hydrogen bound-bound spectrum. Since the number of helium atoms is lower than the number of hydrogen atoms, the effect is not as large, however, additional features appear in the recombination spectrum, e.g., around $\nu\simeq 200\,\GHz$ and $1\,{\rm THz}$. The dominant additional feedback is due to the reprocessing of the $\ion{He}{ii}$ Lyman-$\alpha$ line by $\ion{He}{i}$ continuum absorption. This causes pre-recombination emission of $\ion{He}{i}$ which then feeds back in the $\ion{H}{i}$ Lyman-continuum around $z\simeq 2500-3000$ \citep{Chluba2009c}. For comparison, we also show the case without any feedback at all (dashed/blue) and including primary $\ion{He}{i}$ photon feedback (dotted/violet). The latter case already goes beyond the calculations of \citep{Jose2008}, which neglected the  feedback from the \ion{He}{i} $2^1{\rm S}_0-1^1{\rm S}_0$ two-photon continuum.  

\begin{figure}
\centering 
\includegraphics[width=\columnwidth]{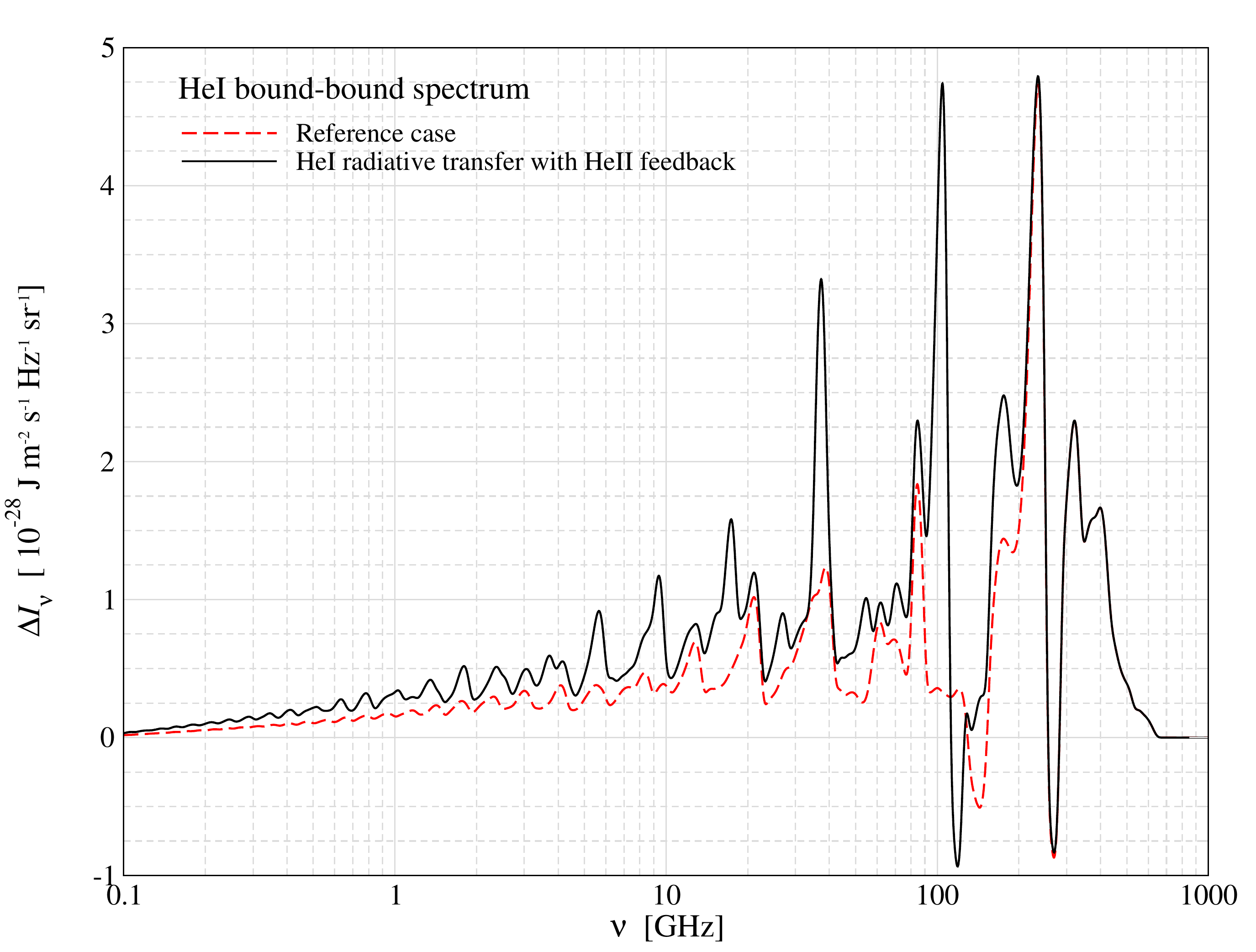}
\caption{Helium feedback corrections to the \ion{He}{i} bound-bound recombination radiation. The conductances were computed for $\nmax=500$. The main features due to \ion{He}{ii} feedback are in good agreement with the 20-shell calculations of \citet{Chluba2009c}.}
\label{fig:HeI_BB_feedback}
\end{figure}
The effect on the $\ion{He}{i}$ recombination spectrum is much more dramatic (see Fig.~\ref{fig:HeI_BB_feedback}). This is because in contrast to the feedback for hydrogen there is roughly one feedback photon per helium atom. The feedback occurs mainly around $z\simeq 4000-5000$ and is dominated by the $\ion{He}{ii}$ Lyman-$\alpha$  \citep{Chluba2009c}. Overall, \ion{He}{ii} feedback enhances the total number of photons emitted by helium itself from $\simeq 6.7\,\gamma/N_{\rm He}$ to $\simeq 8.9 \, \gamma/N_{\rm He}$. We included all free-bound and bound-bound emission of helium atoms in this estimate. Without \ion{He}{ii} feedback, hydrogen emits roughly $5.3\,\gamma/N_{\rm H}$, which with feedback increases to $\simeq 5.4\,\gamma/N_{\rm H}$. In total, $\simeq 6.1\,\gamma/N_{\rm H}$ are emitted in the standard recombination process, with $\simeq 0.7\,\gamma/N_{\rm H}$ contributed by helium alone.

\vspace{-3mm}
\subsection{Derivatives with respect to $Y_{\rm p}$,  $N_{\rm eff}$ and $\Omega_{\rm b}h^2$}
\label{sec:derivs}
As illustrated in \citet{Chluba2008T0}, the CRR directly depends on parameters like the CMB monopole temperature, $T_0$, the baryon density, $\Omega_{\rm b}h^2$, and the expansion rate today, $h$. Similarly, the helium mass fraction, $Y_{\rm p}$, and other parameters affecting the expansion rate, such as the cold dark matter density, $\Omega_{\rm c}$, and effective number of neutrino species, $N_{\rm eff}$, modify the CRR. Although detailed parameter forecasts and consideration of optimal experimental design will be left for future work, here we illustrate how the CRR depends on $\Omega_{\rm b} h^2, Y_{\rm p}$ and $N_{\rm eff}$. Collisions are neglected in the calculation.

\begin{figure}
\centering 
\includegraphics[width=\columnwidth]{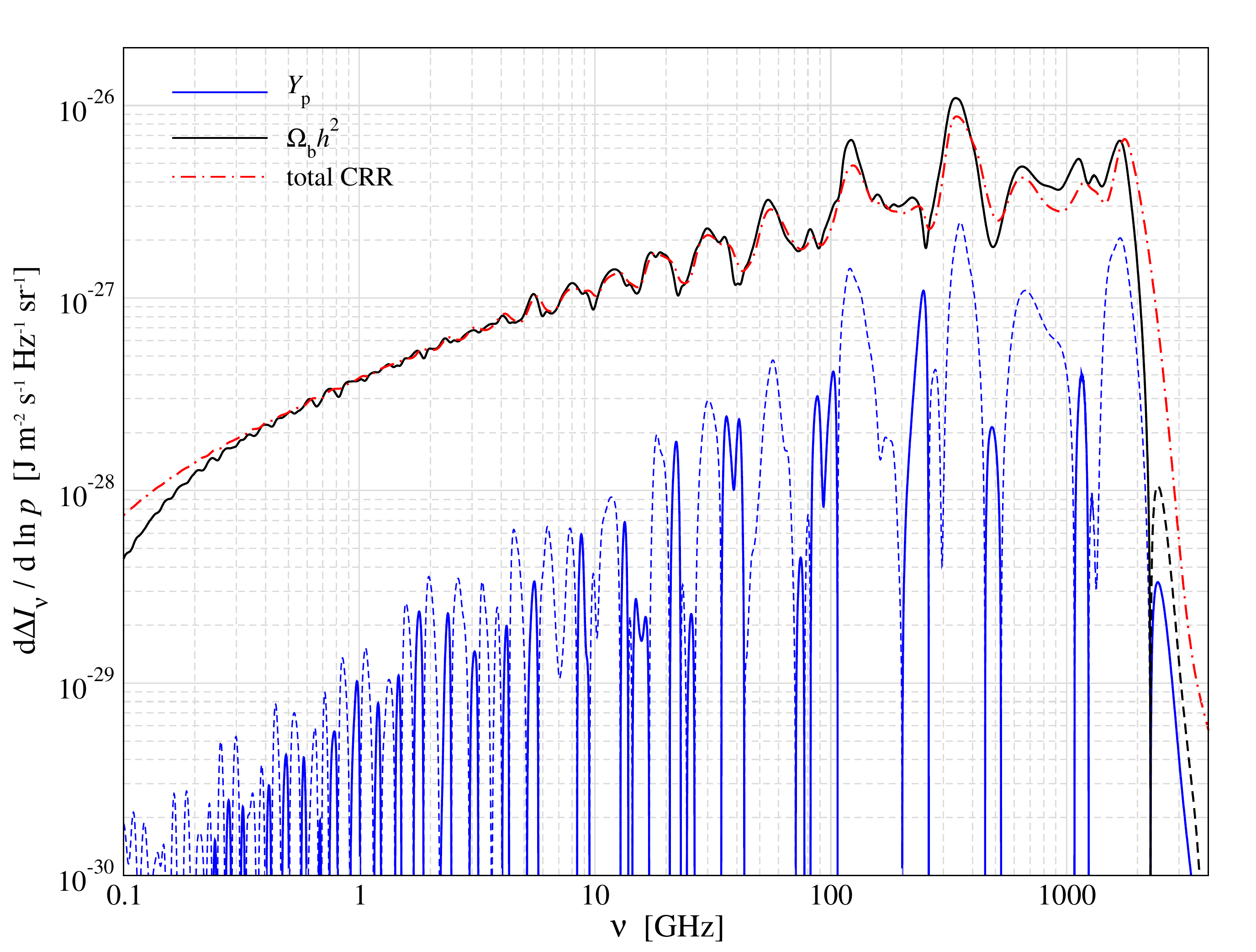}
\\[1mm]
\includegraphics[width=\columnwidth]{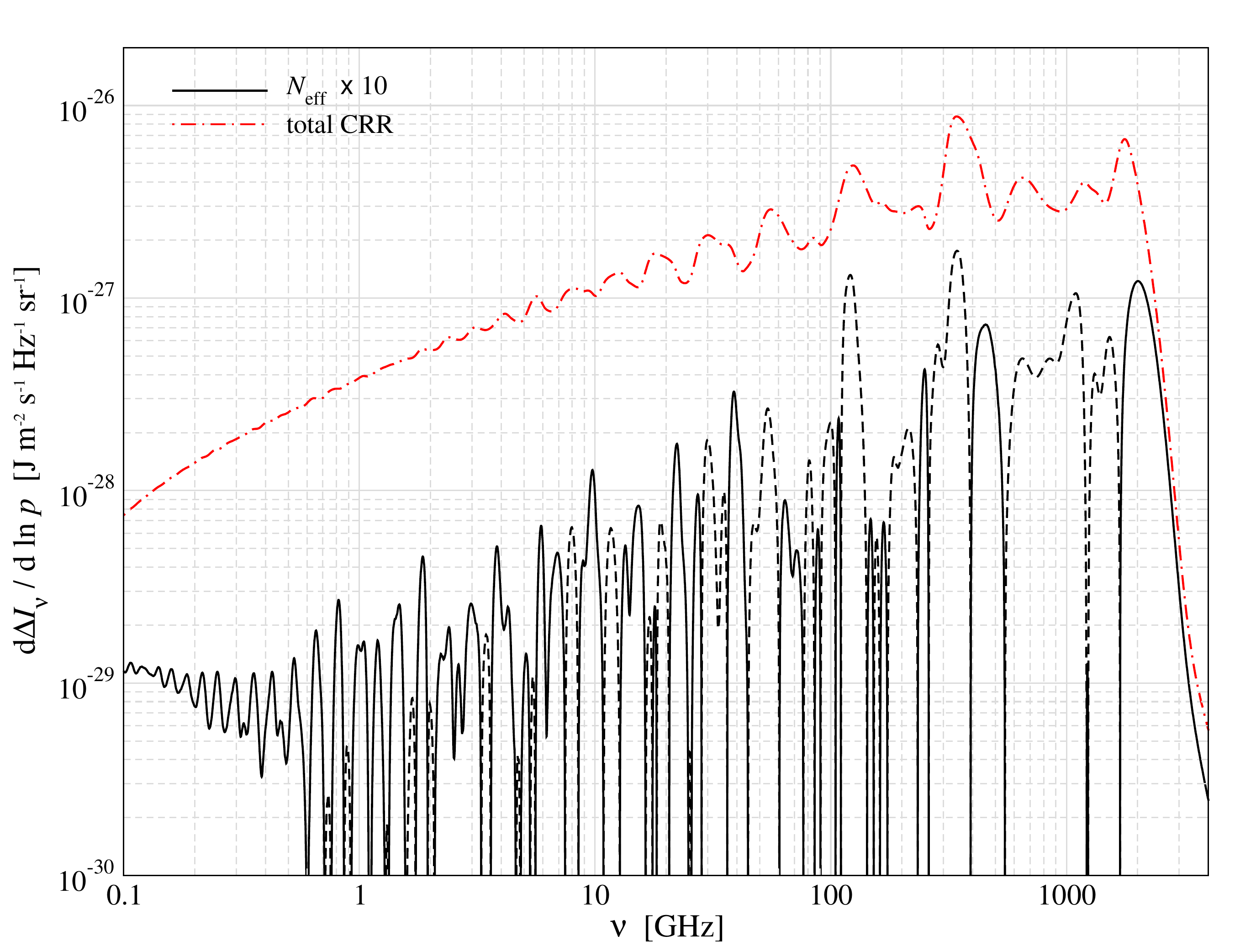}
\caption{Logarithmic derivatives, $\id I_\nu/\id \ln p$, of the total recombination spectrum with respect to $Y_{\rm p}$, $\Omega_{\rm b}h^2$ and $N_{\rm eff}$. The derivatives with respect to $Y_{\rm p}$ and $N_{\rm eff}$ exhibit many features with both positive (solid line) and negative (dashed line) parts. We multiplied the derivative with respect to $N_{\rm eff}$ by a factor of 10 to make it visible on the same scale. For comparison we also show the total CRR itself.}
\label{fig:Deriv_Yp}
\end{figure}

In Fig.~\ref{fig:Deriv_Yp} we show the logarithmic derivatives of the CRR with respect to $\Omega_{\rm b}h^2$ (keeping a flat Universe by varying $\Omega_\Lambda$), $Y_{\rm p}$ and $N_{\rm eff}$. For comparison, we also show the CRR itself. The logarithmic derivative with respect to $\Omega_{\rm b}h^2$ has a similar amplitude and shape as the CRR. This is expected since to leading order the total emission is simply proportional to the number of baryons in the Universe, i.e., $\Delta I_\nu^{\rm CRR}\propto \Omega_{\rm b} h^2$. The differences relative to the CRR are due to changes in the recombination dynamics and at low frequencies the efficiency of free-free absorption.

\begin{figure}
\centering 
\includegraphics[width=\columnwidth]{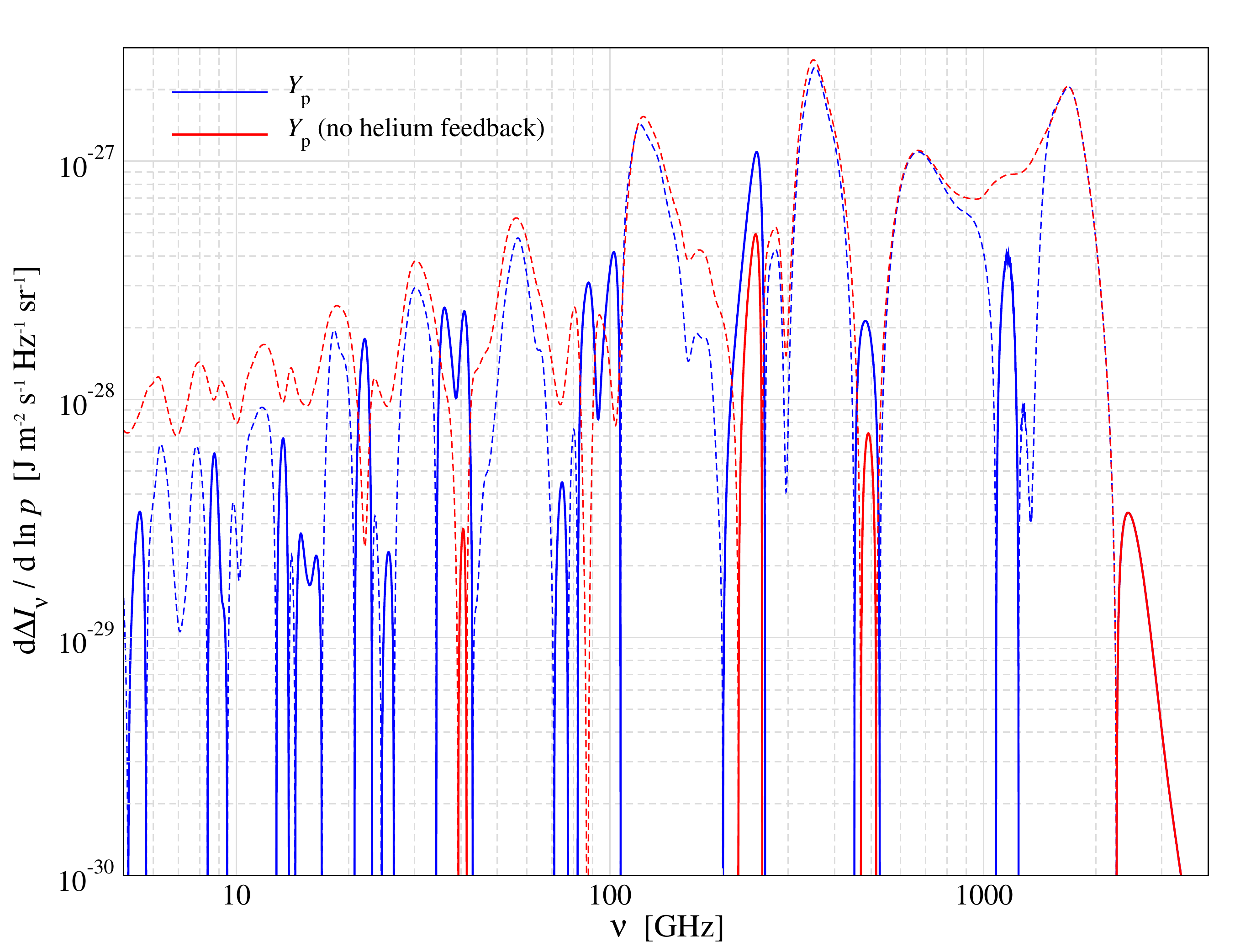}
\caption{Logarithmic derivative, $\id I_\nu/\id \ln p$, of the total recombination spectrum with respect to $Y_{\rm p}$. We compare the case with and without helium feedback. Without helium feedback the derivative has less structure.}
\label{fig:Deriv_Yp_feedback}
\end{figure}

The logarithmic derivatives with respect to $Y_{\rm p}$ and $N_{\rm eff}$ show a much richer structure, with both positive and negative features. The CCR is much less sensitive to changes of $N_{\rm eff}$, since it only enters through the expansion rate. The helium abundance changes both the relative contribution of the two helium spectra to the hydrogen spectrum but also affects the recombination dynamics through feedback and modifications of the recombination rate. Naively, neglecting these dynamical aspects one would simply expect the total CRR to be the sum of the hydrogen and helium templates, which gives fewer features in the derivative (Fig.~\ref{fig:Deriv_Yp_feedback}).
Since the derivatives with respect to $Y_{\rm p}$,  $N_{\rm eff}$ and $\Omega_{\rm b}h^2$ have very different shapes, it is clear that the CRR will help break degeneracies between these parameters. To what extent and with which strategy this may be possible will be studied in detail in a future paper.

\begin{figure}
\centering 
\includegraphics[width=\columnwidth]{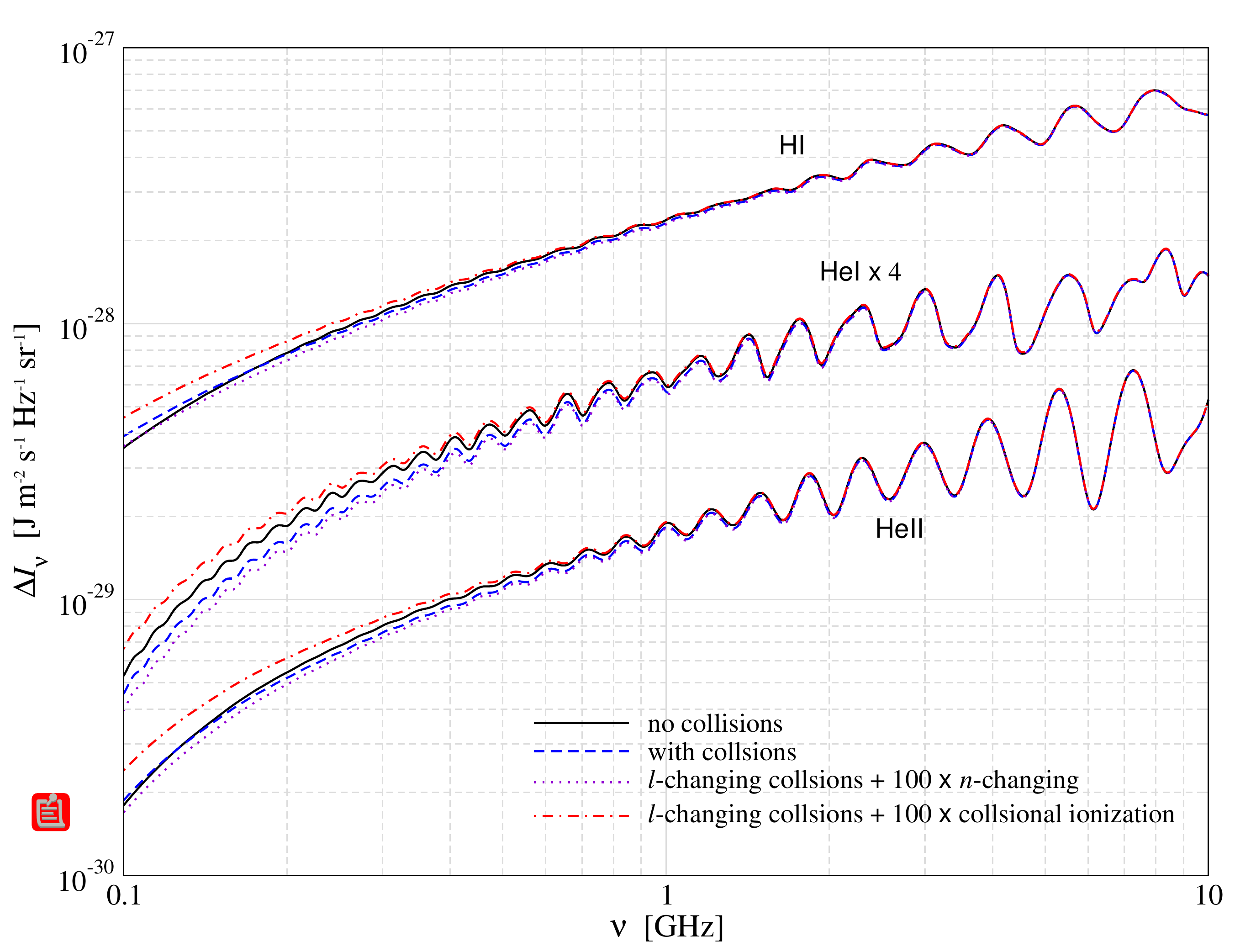}
\caption{Effect of collisions on the recombination radiation for 100 atomic shells. Only $\ell$-changing collisions turn out to be important. Enhancing the rates of $n$-changing collision and collisional ionization rates by a factor of 100 makes their effect  visible at low frequencies. We multiplied the \ion{He}{i} spectrum by 4 to separate it from the \ion{He}{ii} spectrum.}
\label{fig:colls}
\end{figure}

\begin{figure}
\centering 
\includegraphics[width=\columnwidth]{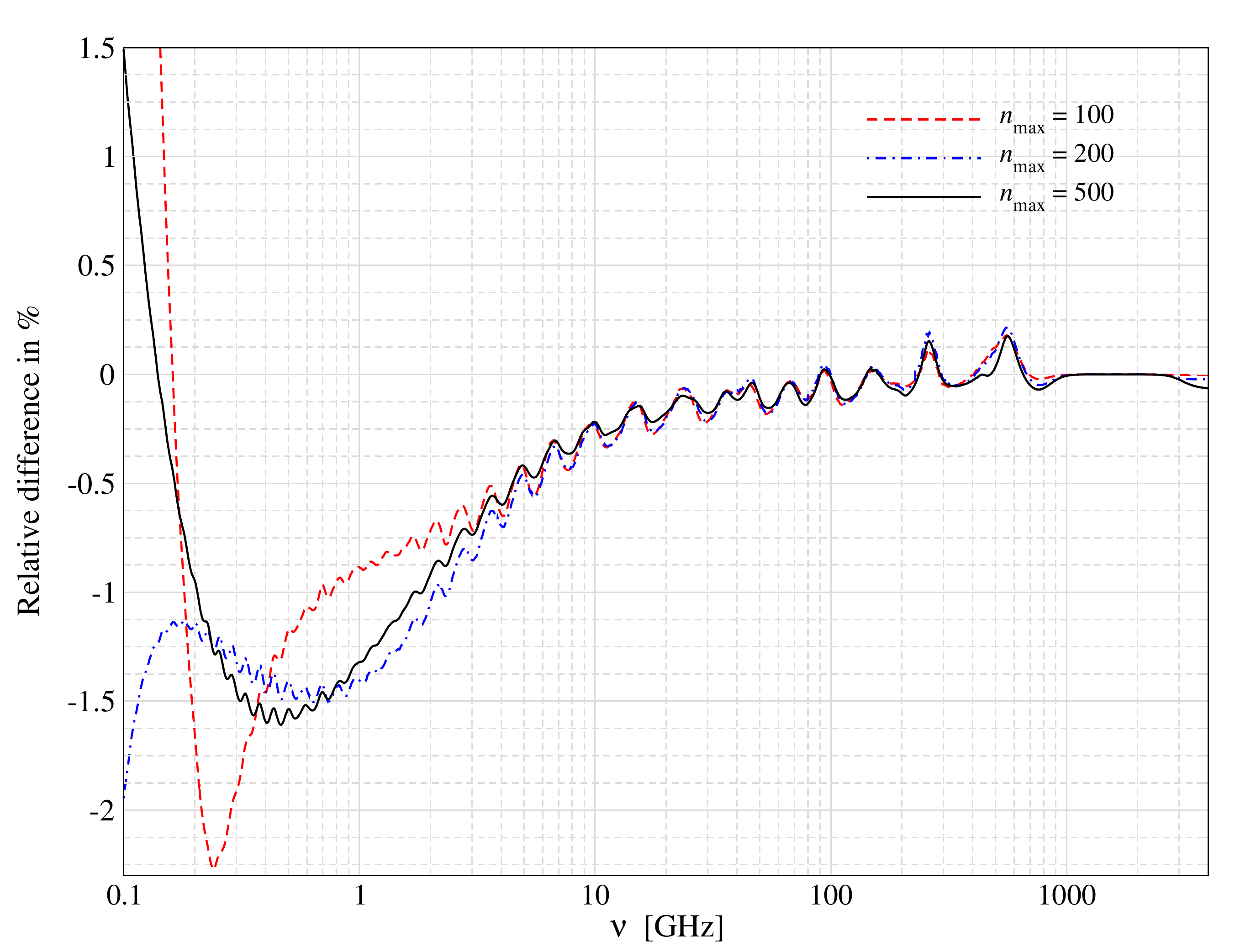}
\caption{Relative change in the total recombination radiation caused by collisions for different number of shells. The spectrum without collisions is chosen as a reference.}
\label{fig:colls_n}
\end{figure}

\vspace{1mm}
\subsection{Effect of collisions}
\label{sec:colls_discussion}
The main effect of $\ell$-changing collisions is to push the angular-momentum substates within a given shell into statistical equilibrium. This enhances the very low-frequency emission ($\nu\lesssim 0.1\,\GHz$) while slightly lowering the emission at higher frequencies. This is because radiative recombinations are more efficient to the low-$\ell$ states ($\ell/n\simeq 0.1-0.2$), which in the absence of collisions are slightly overpopulated in comparison to the high-$\ell$ states. Low-$\ell$ states depopulate via large $\Delta n$ transitions, emitting at high frequencies, whereas transitions among high-$\ell$ states, due to dipole selection rules, are restricted to smaller $\Delta n$ jumps, emitting at lower frequencies. The $\ell$-changing collisions tend to level the low-$\ell$/high-$\ell$ population imbalance and hence enhance the low-frequency emission \citep[see also,][]{Chluba2007}.

Collisional excitations and de-excitations push the relative populations of excited levels closer to Boltzmann equilibrium at the gas temperature, which is very close to the CMB temperature at the relevant redshifts. This therefore lowers the net radiative transition rate, which is proportional to the departure of the level populations from Boltzmann equilibrium with one another.

At low $n$, the populations are closer to Boltzmann equilibrium relative to the lowest ($n = 2$) excited states, while at high $n$ they are closer to Saha equilibrium with the continuum. Collisional ionizations and three-body recombinations drive the highly-excited states even closer to Saha equilibrium with the continuum, steepening the level of departures from Saha equilibrium from low (largest departure) to high $n$ (smallest departure), leaving the low-$n$ states mostly unaffected. 
This has two effects. First, it reduces the net free-bound emission, which is proportional to the population departure from Saha equilibrium. Second, it enhances the bound-bound emission, which depends on the \emph{rate of change} of population departures from equilibrium as a function of $n$.

All these effects are most significant at the low frequencies where the highly excited states radiate. They are also slightly more important for helium than for hydrogen, due to higher densities at the relevant times.  In Fig.~\ref{fig:colls}, we illustrate these effects using 100-shell calculations. We find $n$-changing collisions and collisional ionization to be subdominant, even for the $\ion{He}{ii}$ recombination radiation. Increasing these collision rates by two orders of magnitude makes their effect visible at low frequencies (see figure). The mixing of $\ell$ sub-levels produces a correction to the low-frequency recombination radiation, which for the $\ion{He}{i}$ spectrum becomes slightly more noticeable (in the total CRR of \ion{He}{i} $\simeq 5\%$ as opposed to $\simeq 1.5\%$ for $\ion{H}{i}$ and $\ion{He}{ii}$). However, in all cases the emission at $\nu\gtrsim 5-10\,\GHz$ remains practically unaffected.

In Fig.~\ref{fig:colls_n}, we show the correction to the total CRR caused by collisions. We vary the number of shells in the calculation, to illustrate the changes at low and high frequencies. At low frequencies ($\nu\lesssim 1-2\,\GHz$), the effect reaches percent-level, while at higher frequencies it remains $\lesssim 0.5\%$, nearly independent of $n_{\rm max}$. This shows that collisional processes need to be taken into account for precise computations of the recombination radiation. For baseline calculations we neglect collisions, which implies an uncertainty at percent level for the total CRR.

\begin{figure}
\includegraphics[width=\columnwidth]{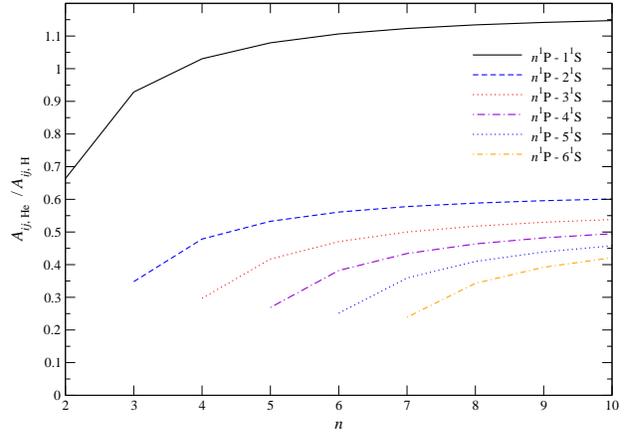}
\centering 
\\[-5mm]
\includegraphics[width=\columnwidth]{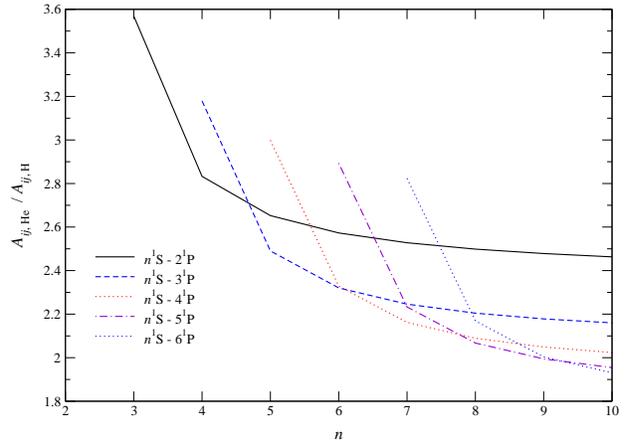}
\caption{Comparison of the \ion{He}{i} singlet transitions rates for the $n$P-$n'$S and $n$S-$n'$P series taken from the multiplet tables of \citet{Drake2007} with the hydrogenic approximation.}
\label{fig:A_sing}
\end{figure}

\vspace{-3mm}
\subsection{Uncertainties due to the neutral helium model}
\label{sec:uncertain_He}
As mentioned above, the atomic model for neutral helium is not as accurate as those of hydrogen and singly-ionized helium. This implies that the neutral helium contributions to the CRR is also uncertain. Here, we illustrate some of the crucial aspects and estimate the level of the uncertainty.

At this point, the level energies are probably the most certain. For $n\leq 10$, we compared the hydrogenic approximation and quantum defect energies with the data from \citet{Drake2007}. The biggest deviations are found for the $n$S and $n$P states, which at $n=10$ depart from the hydrogenic values by $\lesssim 6\%$. The energy values are matched very well by the quantum defect calculation, which is used to extrapolate to higher levels. For $\ell>1$, we find levels with $n\simeq 10$ to depart from the hydrogenic values by $\lesssim \pot{\rm few}{-4}$.

A larger uncertainty is found for the dipole transition rates among low-$\ell$ level involving $n$S, $n$P and $n$D states. In Fig.~\ref{fig:A_sing} we compare the transitions rates from \citet{Drake2007} for the singlet $n$P-$n'$S and $n$S-$n'$P series with the hydrogenic approximation, $A^{\rm He}_{ij}\simeq (\nu^{\rm He}_{ij}/\nu_{\rm ij}^{\rm H})^2\,A^{\rm H}_{ij}$. This approximation works quite well for the $n$P-1S sequence, converging to the hydrogenic approximation at the $\simeq 10\%$ level for $n\simeq 10$. However, for the other cases, the hydrogenic approximation is significantly off, overestimating the transition rates for the $n$P-$n'$S sequence roughly twice and underestimating the $n$S-$n'$P sequence by a factor of $\simeq 2$ around $n\simeq 10$. For the triplet levels, we find a similar mismatch. For singlet transitions involving $n$D states, the hydrogenic approximation becomes accurate at the level of $\simeq 20\%-30\%$. Transition series involving only higher $\ell$ states approach hydrogenic values at the percent level. 
Overall, our analysis implies that low-$\ell$ transitions starting with $n>10$ and ending at $n'\lesssim 10$ are not well approximated using hydrogenic values and an improved neutral helium model including many shells is ultimately needed.

\begin{figure}
\centering 
\includegraphics[width=\columnwidth]{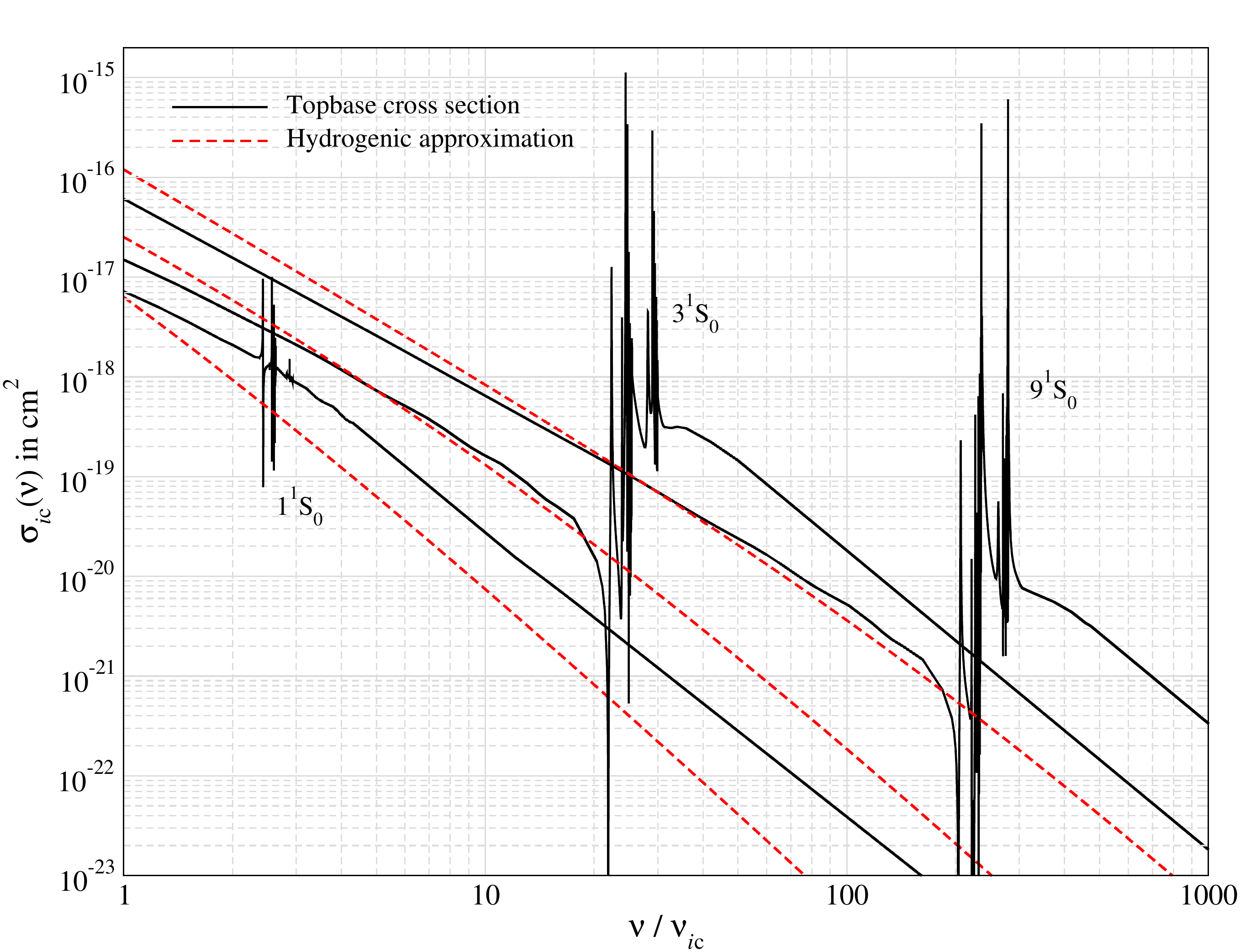}
\caption{Comparison of photonionization cross sections from {\tt Topbase} with simple rescaled hydrogenic approximations. Agreement with the hydrogenic approximation close to the ionization threshold could be achieved by matching the cross section, but even the slopes do not match well.}
\label{fig:He_phys_sigma}
\end{figure}
As mentioned above, the photonionization cross sections are probably the largest source of uncertainty in the neutral helium model. This is because the hydrogenic approximation neglects corrections to the shape of the cross section from the non-hydrogenic character of the wave-functions and due to auto-ionization resonances, which are important for the low-$\ell$ states ($n$S, $n$P and $n$D). 
This is illustrated in Fig.~\ref{fig:He_phys_sigma}, where we compare the {\tt TopBase} cross section with the hydrogenic approximation for some examples. We find that even for $n\simeq 10$, the cross sections for the $n$S and $n$P states show clear non-hydrogenic character already close to the threshold frequency. Although we do not expect auto-ionization resonances to significantly contribute to the total photoionization and recombination rates (which could have an effect on the recombination dynamics) they affect the precise shape of the \ion{He}{i} free-bound continuum. Similarly, for the $n$S and $n$P states we find the slopes of the cross section to be more shallow than the hydrogenic approximation. For the $n$D states, the match is found to be much better already, but auto-ionization resonances are still present even there. 
For the triplet $n$P states, {\tt TopBase} cross sections only exist for $n=2$ and $n=3$. Thus, the shape of the free-bound continuum for $n>3$ is likely affected at a significant level. For $n>9$, we also expect the free-bound continuua to show dependence on the shortcomings of the hydrogenic approximation. A improved quantum mechanical model of neutral helium will therefore eventually be needed for highly accurate predictions of the CRR.

\vspace{-3mm}
\section{Conclusions and future work}
\label{sec:conc}
We carried out the first comprehensive computation of the CRR from hydrogen and helium including all relevant processes for 500-shell atoms. Our computations improve over previous calculations in terms of the total number of included shells and the modeling of feedback processes. For the first time, we include the free-bound emission from helium and model the effect of collisional processes at high redshift. We furthermore overcome the performance-bottleneck of previous treatments using effective multilevel atom and conductance methods. With {\tt CosmoSpec} one calculation of the CRR now only takes $\simeq 15$ seconds on a standard laptop. This opens a way to more detailed forecasts and parameter explorations with the CRR, applications that we will address in a future paper.

The precision of our calculations is currently limited by the atomic model for neutral helium (see Sect.~\ref{sec:uncertain_He} for more detailed discussion) and uncertainties due to collisions. For neutral helium, uncertainties related to the photoionization cross section in particular are expected to exceed the level of 10\%-30\% for the \ion{He}{i} free-bound continuum. This should enter the total CRR at the level of a few percent. Similarly, transition rates, multiplet ratios and radiative/collisional singlet-triplet transitions among excited levels should be considered in more detail. For the total CRR, this implies an uncertainty at the level of a few percent. 
We also show that collisional mixing of $\ell$-substates is noticeable at low frequencies ($\nu\lesssim 2\,\GHz$). At higher frequencies, corrections due to collisions remain at the sub-percent level (Fig.~\ref{fig:colls_n}). 

Our calculations clearly highlight the importance of helium feedback for the detailed shape of the CRR (Fig.~\ref{fig:HI_BB_feedback} and \ref{fig:HeI_BB_feedback}), which introduces additional features into the derivative of the CRR with respect to the helium abundance (Fig.~\ref{fig:Deriv_Yp_feedback}). This process is thus required for a detailed modeling of the dependence of the CRR on the helium abundance, which may be used to extract this parameter from future spectral distortion measurements. The CRR may also help break degeneracies between $N_{\rm eff}$, $Y_{\rm p}$ and $\Omega_{\rm b}$ due to the differing sensitivity of the signal on these parameters. In combination with future CMB anisotropy and large-scale structure measurements \citep{Abazajian2015}, this could deliver additional constraints on neutrino physics.

Our calculations still use a few approximations that could affect the results at the level of $\simeq 1\%$. At this point, we did not separately include the effect of \ion{He}{ii} Lyman-$\beta$. This will change the dynamics of \ion{He}{ii} recombination by Lyman-$n$ feedback \citep{Chluba2007b, Switzer2007I, Yacine2010b} and also modify the interspecies feedback slightly. Similarly, for the \ion{He}{ii} recombination history we neglected detailed radiative transfer effects, stimulated \ion{He}{ii} 2s-1s two-photon emission and Raman processes, all effects that were studied in detail for the hydrogen recombination history \citep{Chluba2006, Kholu2006, Switzer2007III, Chluba2008a, Hirata2008}. 
Corrections due to charge transfer between \ion{He}{ii} and \ion{He}{i} are expected to be small, however, in more detailed computations this processes should be considered. Also, one should treat the direct ionizations from the ground state and continuum escape more carefully.

The CRR probes the Universe at earlier stages than the CMB anisotropies, providing a new window for complementary constraints on non-standard physics. For instance, variations of fundamental constants, such as the fine-structure constant or electron-to-proton mass ratio, would shift the position and relative amplitudes of the recombination lines. 
We highlight that the effect of variations of $\alpha$ and $\me$ can be treated self-consistently using the effective rate and conductance method (see Appendix~\ref{app:rescalings}). We find the effective recombination and photoionization rates to scale as $\alpha_{{\rm c} i} \propto \alpha^2 \me^{-2}$ and $\beta_{i\rm c}\propto \alpha^5 \me$, respectively. These effects have been incorporated by {\tt HyRec} for a while\footnote{See supplementary material on the {\tt HyRec} webpage.}, but they differ from previously adopted scalings \citep[e.g.,][]{Kaplinghat1999, Scoccola2009}. The differences may change the constraints derived from {\it Planck} data \citep{Planck2015var_alp}.

Another interesting effect is the extra ionization and excitation from annihilating or decaying dark matter particles, which could significantly enhance the CRR. It may therefore be possible to derive interesting upper limits on dark matter properties with future spectral distortion measurements, even in the absence of a detection. Tests of spatial variations in the composition of the Universe, for example, due to non-standard BBN models or cosmic bubble collisions \citep{Dai2013, Silk2014Sci, Chary2015} may be another line of research. We look forward to exploring these directions in the future.

\vspace{-0mm}
\small
\section*{Acknowledgments}
JC is supported by the Royal Society as a Royal Society University Research Fellow at the University of Cambridge, UK. YAH is supported by NSF Grant No. 0244990,
NASA NNX15AB18G, the John Templeton Foundation, and the Simons Foundation.
Use of the GPC supercomputer at the SciNet HPC Consortium is acknowledged. SciNet is funded by: the Canada Foundation for Innovation under the auspices of Compute Canada; the Government of Ontario; Ontario Research Fund - Research Excellence; and the University of Toronto.

\small 

\begin{appendix}

\vspace{-4mm}
\section{Improved $J$-resolved transitions rates from hydrogenic values}
\label{app:transitions}
In the helium model of \citet{Jose2008}, transitions among triplet states with $n\leq 10$ which are not contained in the multiplet tables of \citet{Drake2007} are filled with hydrogenic values using the approximation
\beal
\label{eq:A_approx}
A^{\rm He}_{n L J\rightarrow n' L' J'}&\approx\frac{(2J'+1)}{(2L'+1)(2S'+1)}\,A^{\rm H}_{n L \rightarrow n' L'}.
\end{align}
This approximation assumes that the partial rates for individual sub-levels within a multiplet are the same and that the total transition rate averaged over all $J$ and $J'$ is given by the hydrogenic (non-$J$-resolved) values. This can be readily confirmed by computing the $J$-averaged transition rate\footnote{This expression assumes statistical equilibrium values for the populations of level $i=\{nLJ\}$ according to $N_{nLJ}\approx \frac{(2J+1)}{(2L+1)(2S+1)} \bar{N}_{nL}$, where $\bar{N}_{nL}$ is the $J$-averaged population.}
\beal
\label{eq:A_av}
A^{\rm He}_{n L\rightarrow n' L'}&=\sum_{J J'} \frac{(2J+1)}{(2L+1)(2S+1)}\,A^{\rm He}_{n L J\rightarrow n' L' J'}\approx A^{\rm H}_{n L \rightarrow n' L'}
\end{align}
after inserting Eq.~\eqref{eq:A_approx}.
From the atomic physics point of view, this approximation is incorrect. It is well-known \citep[e.g.,][]{Goldberg1935, Condon1963} that the transitions within a given multiplet differ in strength and that their relative ratio deviates by much more than expressed with Eq.~\eqref{eq:A_approx}. Using the relation $\mathcal{S}_{ij}\propto \lambda^3 g_i \, A_{ij}$, between the line strength $\mathcal{S}_{ij}$ and transition rate $A_{ij}$ together with the sum rule \citep[e.g., see][]{Edmonds1960}
\beal
\label{eq:S_av}
\mathcal{S}_{n L J\rightarrow n' L' J'}=(2J+1)(2J'+1)\left|\WignerSixJ{L}{J}{S}{J'}{L'}{1}\right|^2\mathcal{S}_{n L\rightarrow n' L'},
\end{align}
where $\WignerSixJ{a}{b}{c}{d}{e}{f}$ denotes the Wigner-6$J$-symbol, one can show that the transition rates within a multiplet should scale like
\beal
\label{eq:A_av_approx}
A^{\rm He}_{n L J\rightarrow n' L' J'}\approx (2L+1)(2J'+1)\left|\WignerSixJ{L}{J}{S}{J'}{L'}{1}\right|^2 A^{\rm H}_{n L\rightarrow n' L'}.
\end{align}
For $\Delta L=-1$, we therefore have the three cases
\bsub
\beal
\label{eq:A_av_approx_ratios_I}
r_{J-1}&= \frac{(L+J+2)(L+J+1)(L+J-1)(L+J-2)}{4J(2J+1)\,L\,(2L-1)}
\\
r_J&= \frac{(2J+1)(J-L-1)(L-J-2)}{(J+1)(L+J+1)(L+J-2)}\,r_{J-1}
\\
r_{J+1}&= \frac{J(J-L)(L-J-3)}{(2J+1)(L+J-1)(L+J+2)}\,r_{J}
\end{align}
\esub
for the line ratios $r_k=A^{\rm He}_{n L J\rightarrow n' L-1 k}/A^{\rm H}_{n L\rightarrow n' L-1}$. Within these multiplets, the strongest line is the one for $\Delta J=-1$, which for $L\gg 1$ roughly equals the hydrogenic value, $r_{J-1}\approx 1$. For large $L$, the intensity of the second line ($\Delta J=0$) drops like $r_J\approx L^{-2}$, while for the third line we have $r_{J+1}\approx 4L^{-4}$. Clearly, this is a very different behavior than obtained with Eq.~\eqref{eq:A_approx}.
For $\Delta L=+1$, we similarly find the line ratios
\bsub
\beal
\label{eq:A_av_approx_ratios_II}
r_{J-1}&= \frac{(J+1)(J-L-3)(L-J)}{(2J+1)(L+J)(L+J+3)}\,r_J
\\
r_J&= \frac{(2J+1)(L-J-1)(J-L-2)}{J(L+J+1)(L+J+4)}\,r_{J+1}
\\
r_{J+1}&= \frac{(L+J+4)(L+J+3)(L+J+1)(L+J)}{4(2L+3)(L+1)(2J+1)(J+1)},
\end{align}
\esub
where now the strongest line in the multiplet is the one for $\Delta J=+1$. 

When considering transitions from levels with $n>10$ (non-$J$-resolved) to levels with $n'\leq 10$ ($J$-resolved), we average Eq.~\eqref{eq:A_av_approx} over the initial level $J$, giving us the approximation
\beal
\label{eq:A_av_approx_J}
A^{\rm He}_{n L\rightarrow n' L' J'}\approx \frac{(2J'+1)}{(2L'+1)(2S'+1)}\,A^{\rm H}_{n L\rightarrow n' L'},
\end{align}
which also follows from Eq.~\eqref{eq:A_approx}. Notice that to obtain the hydrogenic approximation we also scale by the average transition frequency ratio to the third power. This usually is a small correction.

We find that the changes to the recombination history and spectrum caused by the above improvements are small ($\lesssim 0.1\%$). The main reason is that only a few transitions among the high-$L$ states at $n\leq 10$ are missing from the multiplet tables of \citet{Drake2007}. In addition, for transitions between levels with $n>10$ (non-$J$-resolved) to levels with $n'\leq 10$ ($J$-resolved) the approximation of \citet{Jose2008} already represented the correct value, as shown in Eq.~\eqref{eq:A_av_approx_J}. This explains why the overall correction remains small.

\vspace{6mm}
\subsection{$J$-resolved photoionization cross sections}
\label{app:sigma}
The total photoionization cross section from level $n L J$ is the sum of the photoionization cross sections to continuum states (label ``c", with an implicit integration over final electron momentum) with angular momentum quantum numbers $L' J'$:
\beq
\sigma_{n L J} = \sum_{L' J'} \sigma_{n L J\rightarrow \textrm{c} L' J'}, 
\eeq
where the spin quantum number $S$ is implicit and held fixed throughout. Each angular-momentum-resolved transition is proportional to the line strength $\mathcal{S}_{n L J, \textrm{c} L' J'}$ \citep{Condon1963}:
\beq
\sigma_{n L J\rightarrow \textrm{c} L' J'} \propto \frac{\mathcal{S}_{n L J, \textrm{c} L' J'}}{2 J +1}.
\eeq
Summing over final states, and using the sum rule \citep{Edmonds1960}
\beq
\sum_{J'} \mathcal{S}_{n L J, \textrm{c} L' J'} = \frac{2 J +1}{2 L +1} \mathcal{S}_{n L, \textrm{c} L'},
\eeq
we arrive at 
\beq
\sigma_{n L J} \propto \frac1{2 L + 1} \sum_{L'} \mathcal{S}_{n L, \textrm{c} L'},
\eeq
which is independent of $J$. 

\vspace{5mm}
\section{Rescaling of the effective rates for hydrogenic ions}
\label{app:rescalings}
Noting that the dipole transition rates, $A_{i\rightarrow j}$ (between levels $i$ and $j$, where $i$ is the upper level) and photoionization cross sections\footnote{For the photoionization cross section, this can most easily be seen using Kramers' formula \citep[see][]{Karzas1961}, 
\beal
\sigma^{\rm K}_{n\ell\rightarrow \rm c}=\frac{2^4}{3\sqrt{3}\,n}\,\frac{e^2}{\me c}\frac{\nu_{i \rightarrow {\rm c}}^2}{\nu^3}\propto \alpha\, \me^{-1}\,\nu_{i \rightarrow {\rm c}}^2 \,\nu^{-3}\propto  \alpha\,a_0^2\,\nu_{i \rightarrow {\rm c}}^3 \,\nu^{-3}\,Z^{-2}.
\end{align}
For the transition rate, it directly follows from Fermi's Golden rule.
}, $\sigma_{i\rightarrow \rm c}$, scale as 
\beal
A_{i\rightarrow j} 
&\propto \sigma_{i\rightarrow \rm c} \nu^2 \id \nu \propto \alpha\,\nu_{i \rightarrow j/{\rm c}}^3\,a_0^2 \,Z^{-2}= \alpha^5\,\me \,Z^4,
\end{align}
we find that the radiative rate coefficients scale as  \citep[compare,][]{Yacine2010c}
\bsub
\beal
R_{i\rightarrow j} &= A_{i\rightarrow j}[1+n_{\nu_{ij}}(\Tg)] = \alpha^5\,\me \,Z^4 
F\left(\frac{k\Tg}{h\nu_{ij}}\right)
\\[0mm]
\beta_{i\rightarrow {\rm c}} &= \frac{8\pi}{c^2}\int_{\nu_{i\rm c}}^\infty  \nu^2 \,\sigma_{i\rightarrow \rm c}(\nu) \,n_\nu(\Tg) \id \nu
= \alpha^5\,\me\,Z^4 B\left(\frac{k\Tg}{h\nu_{i\rm c}}\right)
\\[0mm]
\alpha_{{\rm c}\rightarrow i} &= \frac{8\pi}{c^2}\,f_{i}^{\rm eq}(\Te)\int_{\nu_{i\rm c}}^\infty  \nu^2 \,\sigma_{i\rightarrow \rm c}(\nu) \,[1+n_\nu(\Tg)] \,\expf{-h\nu/k\Te}\id \nu 
\nonumber\\
&= \frac{\alpha^2\,Z}{\me^2} A\left(\frac{k\Tg}{h\nu_{i\rm c}},\frac{k\Te}{h\nu_{i\rm c}}\right)
\\[0mm]
f_{i}^{\rm eq}(\Te)&=\frac{g_i}{2 g_{\rm c}}\frac{h^3\,\expf{h\nu_{i\rm c}/k\Te}}{(2\pi \me k\Te)^{3/2}}.
\end{align}
\esub
Here, $\alpha$ is the fine structure constant; $Z$ the charge of the hydrogenic ion; $\me$ its reduced mass; $a_0\simeq \alpha^{-1} \me^{-1}$ the Bohr radius; $\nu_{ij}\simeq \nu_{i\rm c} \simeq \alpha^2 \me\,Z^2$ the bound-bound / bound-free transition frequency, respectively; $n_{\nu}(\Tg)=1/(\expf{h\nu/k\Tg}-1)$ the blackbody photon occupation number; $g_i$ and $g_{\rm c}$ the statistical weights of the level $i$ and the continuum particle. 

The inverse dipole transitions rate, $R_{j\rightarrow i}$, is simply given by $R_{j\rightarrow i}=\frac{g_i}{g_j} R_{i\rightarrow j}\,\expf{-h\nu_{ij}/k\Tg}$. 
Because the probabilities of an electron reaching a specific level $i$ from another level $j$ within the excited levels just depends on ratios and sums of $R_{i\rightarrow j}$ and $\beta_{i\rightarrow {\rm c}}$, the effective rate coefficients $\mathcal{A}_i(\Tg, \Te)$, $\mathcal{B}_i(\Tg)$ and $\mathcal{R}_{i\rightarrow j}(\Tg)$ \citep[Eq. 17, 18 and 19 of][]{Yacine2010c} for any hydrogenic ion can be obtained from those of hydrogen. Defining the factors $f_\mathcal{A}=\alpha^2\,Z\,\mu^2_{\rm H}/[\alpha^2_{\rm ref}\,\mu^2_Z]$, $f_\mathcal{B}=\alpha^5\,Z^4\,\mu_{Z}/[\alpha^5_{\rm ref}\,\mu_{\rm H}]$ and $f_T=\alpha^2_{\rm ref}\,\mu_{\rm H}/[\alpha^2 Z^2\,\mu_Z]$, we find
\bsub
\beal
\mathcal{A}_i(\Tg, \Te)&=f_\mathcal{A}\,
\mathcal{A}^{\rm H}_i\left(f_T\Tg, f_T\Te\right)
\\
\mathcal{B}_i(\Tg)&=f_\mathcal{B}\,
\mathcal{B}^{\rm H}_i\left(f_T \Tg\right)
\\
\mathcal{R}_{i\rightarrow j}(\Tg)&=f_\mathcal{B}\,
\mathcal{R}^{\rm H}_{i\rightarrow j}\left(f_T\Tg\right),
\end{align}
\esub
where $\mu_{\rm H}$ and $\mu_Z$ are the reduced masses of hydrogen and the hydrogenic ion with charge $Z$ in units of the electron rest mass. We also included the explicit dependence on $\alpha$, where $\alpha_{\rm ref}\approx 1/137$ is the valued used for the effective rate computation.

The effective conductances of hydrogenic helium can be similarly obtained with
\bsub
\beal
\mathcal{G}^{\rm e}_{n' n}(\Tg)&=f_\mathcal{B}\,
\mathcal{G}^{\rm e, H}_{n' n}\left(f_T\Tg \right)
\\
\mathcal{G}^i_{n' n}(\Tg)&=f_\mathcal{B}\,
\mathcal{G}^{i, \rm H}_{n' n}\left(f_T\Tg \right)
\\
\frac{\id \mathcal{G}^i_{\rm fb}(\Tg)}{\id \nu}&=f_\mathcal{B}\,
\frac{\id \mathcal{G}^{i, \rm H}_{\rm fb}}{\id \nu}\left(f_T\Tg \right)
\end{align}
\esub
using those of hydrogen (superscript 'H') with the notation of \citet{Yacine2013RecSpec} for the conductances. In addition, we need to shift the energies of each frequency bin in the conductance table. These simple scalings are no longer exactly valid when collisions are included (Sect.~\ref{sec:colls}). They also neglect smaller corrections due to the Lamb-shift, which depend on $\alpha$ as well.

\vspace{-4mm}
\section{Classical derivation of $\ell$-changing collisional transition rates}
\label{app:coll}
In this Appendix, we give a simple derivation of the classical result of \cite{Vrinceanu2012} in the Born approximation. We recover the leading term of their classical result up to corrections of order $1/n \ll 1$.

Let us consider an electron moving in a Keplerian orbit around a nucleus with charge $Z_{\rm nuc}$, and subject to a
perturbing force per unit mass $\bs{f}(\bs{r}, t)$ acting on the electron. We have $Z_{\rm nuc} = 2$ for singly-ionized helium and 1 for hydrogen and neutral helium, for which we approximate the helium nucleus and inner electron as a point charge, and the orbiting electron is the outermost, excited one.

The Gaussian perturbation equation \citep[see e.g.][]{Roy1982} for the osculating orbital elements $c_i$ are
\beq
\frac{\id c_i}{\id t} = \frac{\partial c_i}{\partial \bs{\varv}} \cdot \bs{f}.
\eeq
We emphasize that these equations are exact.
We define $\kappa \equiv Z_{\rm nuc} q_{\rm e}^2/\me$, where $q_{\rm e}$ is the elementary charge and $\me$ is the reduced mass of the electron-nucleus system. The energy per unit mass is $\epsilon =
\frac12 \varv^2 - \kappa/r$, and evolves under the influence of the
perturbing force according to 
\beq
\frac{\id\epsilon}{\id t} = \bs{\varv \cdot f}, 
\eeq
and the angular momentum per unit mass $\bs{L} \equiv \bs{r} \times \bs{\varv}$
evolves according to 
\beq
\frac{\id\bs{L}}{\id t} = \bs{r} \times \bs{f}.
\eeq
We now make two assumptions: \\
$(i)$ We assume that the perturbing force is nearly constant accross
the orbit of the electron. For the problem of interest ($\bs{f}$
generated by a passing ion), this is equivalent to
the electric dipole approximation, where we
assume that the dimensions of the orbit are much smaller than the
impact parameter, $b$.\\
$(ii)$ We assume that the force varies on a
timescale long compared to the orbital period. This is the orbital
adiabatic approximation, and it requires that $\varv/b \ll \nu_{n, n\pm1}$, where $\nu_{n, n\pm1}$ is
the transition frequency from $n$ to the neighboring energy states.

With these assumptions, one can average the rate of change of orbital elements over one orbit, and obtain, first, that the energy $\epsilon$ is conserved, and second, that the secular rate of change of angular momentum is
\beq
\Bigg{\langle}\frac{\id\bs{L}}{\id t}\Bigg{\rangle} = \langle\bs{r}\rangle \times \bs{f},
\eeq
where $\langle\bs{r}\rangle$ is the time-average of the position
vector over one orbit. Note that this reduces to the standard formula
for a torque on an electric dipole moment when $\me\bs{f} =  q_{\rm e} \bs{E}$. 

The time-averaged position vector for a Keplerian orbit can be shown to be $\langle \bs{r} \rangle =  - (3/2) a \bs{e}$, where $a$ is the semi-major axis and $\bs{e}$ is the eccentricity vector, pointing from the orbital focus to percienter, and having magnitude $e < 1$, the orbital eccentricity. The secular rate of change of angular momentum is therefore
\beq
\Bigg{\langle}\frac{\id\bs{L}}{\id t}\Bigg{\rangle} = \frac32 a \bs{f} \times
\bs{e}.
\eeq
One can derive a similar expression for the rate of change of the eccentricity vector itself,
\beq
\Bigg{\langle}\frac{\id\bs{e}}{\id t}\Bigg{\rangle} = \frac3{2\kappa} \bs{f} \times \bs{L},
\eeq
though we will not need it in the Born approximation.


We now specify to the perturbing force created by a passing ion of
charge $Z_{\rm ion} q_{\rm e} > 0$, moving on a trajectory $\bs{R}(t)$:
\beq
\bs{f} = \frac{Z_{\rm ion} q_{\rm e}^2}{\me} \frac{\hat{R}}{R^2}.
\eeq 
The specific angular momentum $L_{\rm ion}$ of the ion-nucleus system is conserved in the collision (up to corrections of order $\hbar$). Defining $\Phi$ as the polar angle of the passing ion, which starts at $\Phi = 0$, we have $L_{\rm ion} = R^2 \dot{\Phi} = b \varv$, where $b$ is the impact parameter and $\varv$ the velocity of the ion at infinity. We may therefore rewrite the perturbing force as 
\beq
\bs{f} = \frac{Z_{\rm ion} q_{\rm e}^2}{\me b \varv} \dot{\Phi} \left[ \cos \Phi ~\hat{x} + \sin \Phi~ \hat{y} \right],
\eeq
where $\hat{x}, \hat{y}$ form a fixed orthonormal basis in the plane of the ion's orbit, with $\hat{x}$ pointing in the incoming direction of the ion.

We now define the dimensionless angular momentum 
\beq
\bs{l} \equiv \frac{\bs{L}}{\sqrt{\kappa a}},
\eeq
whose magnitude is $l = \sqrt{1 - e^2}$. In terms of quantum numbers for a hydrogenic atom, $l = \ell / n$. The secular equation can be rewritten as
\beq
\frac{\id\bs{l}}{\id \Phi} = \alpha\left[\cos \Phi ~\hat{x}
  + \sin \Phi ~\hat{y}\right] \times \bs{e}, \label{eq:dl_dPhi}
\eeq
where 
\beq
\alpha \equiv \frac{3}{2} \sqrt{\frac{a}{\kappa}} \frac{Z_{\rm ion} q_{\rm e}^2}{\me b \varv}.
\eeq
So far we have not made any approximation besides the orbital
adiabatic and electric dipole approximations. We now make the Born
approximation, i.e.~we assume that the perturbation is small enough
that $\bs{e}$ can be approximated as constant in the above equation. This
is equivalent to a first-order expansion in the small parameter $\alpha \ll 1$. We also assume that the ion's trajectory can be approximated as a straight line, with final polar angle $\Phi_f = \pi$. This is accurate for neutral hydrogen and helium (up to small deflections due to the induced polarization of the atom by the passing ion). It is not necessarily so for singly-ionized helium, in which case the ion's orbit is a hyperbola. We shall argue later on that for the conditions relevant to this problem the deflection angle is very small and the straight line approximation should be very accurate, even for singly-ionized helium.

Integrating Eq.~\eqref{eq:dl_dPhi} from $\Phi = 0$ to $\pi$, we get
\beq
\Delta \bs{l} = 2 \alpha e ~\hat{y}.
\eeq
The change of the magnitude of $\bs{l}$ is therefore to first order in $\alpha$, 
\beq
\Delta l = 2 \alpha e ~ \hat{y} \cdot \hat{l}.
\eeq
When considering all orientations of the system, the scalar product $\hat{y} \cdot \hat{l}$ is uniformly distributed between -1 and 1. Therefore, at fixed $b$ and $\varv$, the differential probability to change the angular momentum by $\Delta l$ is 
\beq
\frac{\id P}{\id \Delta l}  = 
    \begin{cases}
         \frac1{4 \alpha e} & \ \text{if} \ |\Delta l| \leq 2 \alpha e,\\
         0 & \text{otherwise}.
    \end{cases} \label{eq:dPdl}
\eeq
The differential collision rate is then obtained from integrating over impact parameters and velocities:
\beq
\frac{\id C}{\id\Delta l}  = N_{\rm ion} \int \id^3 \varv ~\varv f_{\rm ion}(\varv) \int 2 \pi b \id b
\frac{\id P}{\id \Delta l},
\eeq
where $f_{\rm ion}(\varv)$ is the Maxwell-Boltzmann probability distribution for the velocity $\varv$ of the ion relative to the target, 
\beq
f_{\rm ion}(\varv) \equiv \left(\frac{M}{2 \pi kT}\right)^{3/2} \exp \left[ - \frac{M \varv^2}{2 kT} \right],
\eeq
where $M$ is the reduced mass of the ion-target system. Changing integration variables from $b$ to $\alpha$, we arrive at
\beq
\frac{\id C}{\id \Delta l}  = N_{\rm ion} 2 \pi \left(\frac{3}{2}\sqrt{\frac{a}{\kappa}} \frac{Z_{\rm ion}
q_{\rm e}^2}{\me}\right)^2 \int \frac{\id^3 \varv}{\varv} f_{\rm ion}(\varv)  \int \frac{\id P}{\id \Delta l} \frac{\id \alpha}{\alpha^3}.
\eeq
For any given $\Delta l$, the classical transition probability is
non-vanishing only for $\alpha \geq |\Delta l|/(2 e)$, and
we therefore obtain
\beq
\int \frac{\id P}{\id \Delta l} \frac{\id\alpha}{\alpha^3} = \frac1{4 e} \int_{|\Delta l|/(2 e)} \frac{\id\alpha}{\alpha^4} = \frac23 \frac{1 - l^2}{|\Delta l|^3}, \label{eq:int-alpha}
\eeq
where we used $l^2 + e^2 = 1$. This is the first term in the semi-classical expansion of \cite{Vrinceanu2012} [their Eq.~(7)], the next terms being suppressed by higher orders in $|\Delta l| \ll 1$. Integrating over velocities and inserting the value of $\kappa$, we obtain
\beq
\frac{\id C}{\id \Delta l} = N_{\rm ion} \frac{3  a}{Z_{\rm nuc}} \frac{(Z_{\rm ion} q_{\rm e})^2}{\me} \sqrt{\frac{2\pi
    M}{k T}} \frac{1 - l^2}{|\Delta l|^3}.
\eeq
Finally, substituting $a = n^2 a_0/Z_{\rm nuc}$, where the Bohr radius is
\beq
a_0 = \frac{\hbar^2}{\me q_{\rm e}^2},
\eeq
and replacing $l = \ell / n$, we get our final expression,
\beq
\frac{\id C}{\id\Delta \ell} = N_{\rm ion} \frac{Z_{\rm ion}^2}{Z_{\rm nuc}^2} \,3 n^4 \frac{\hbar^2}{\me^2} \sqrt{\frac{2 \pi
    M}{k T}} \frac{1 - (\ell/n)^2}{|\Delta \ell|^3}.
\eeq
We emphasize that this is a \emph{differential} transition rate, since classically $\ell$ is a continuous number. 

To get the expression of \cite{Vrinceanu2012}, we approximate
\beq
C_{n \ell \rightarrow n \ell'} \approx \frac{\id C}{\id \Delta \ell} (\Delta \ell = \ell' - \ell).
\eeq
This expression is of course expected to be only accurate within order unity factors. As shown by \citet{Vrinceanu2012}, their expression reproduces the full quantum calculations for $\Delta \ell>1$ extremely well.

\vspace{-0mm}
\subsection{A note on hyperbolic trajectories}

We have assumed that the trajectory of the colliding ion relative to the target nucleus is a straight line. In the case of singly-ionized helium, the target is charged and the orbit is in fact a hyperbola. One can show in this case that the final polar angle $\Phi_f$ is such that 
\beq
\tan(\Phi_f/2) = \frac{M b \varv^2}{Z_{\rm ion} q_{\rm e}^2} = \frac3{2 \alpha} \frac{M}{\me} \frac{\varv}{\sqrt{2 |\epsilon|}}.
\eeq 
Now taking $\varv \sim \sqrt{k T/M}$ and $|\epsilon| = E_{\rm I}/(n^2 \me)$, where $E_{\rm I}$ is the ionization energy, we obtain
\beq
\tan(\Phi_f/2) \sim \alpha^{-1} n \sqrt{\frac{M}{\me}} \sqrt{\frac{k T}{E_{\rm I}}}.
\eeq
For a straight line trajectory, we saw previously that the collision rate is a steeply decreasing function of $\alpha$ [see Eq.~\eqref{eq:int-alpha}], and is dominated by values of $\alpha$ near the threshold value $|\Delta l|/(2 e) \sim 1/n$. We therefore get
\beq
\tan(\Phi_f/2) \sim n^2 \sqrt{\frac{M}{\me}} \sqrt{\frac{k T}{E_{\rm I}}}.
\eeq
The recombination from doubly ionized to singly-ionized helium takes place at $z \sim 6000$ corresponding to $k T \sim 1.6$ eV. The ionization energy of hydrogen-like helium is 54 eV. For a colliding proton $M/\me \sim 1500$. We therefore get
\beq
\tan(\Phi_f/2) \sim 7 ~n^2.
\eeq
Since we are interested mostly in highly excited states $n \gg 1$ we see that the deflection angle $\pi - \Phi_f \sim 1/n^2 \ll 1$ and the straight-line approximation is therefore very good even for singly-ionized helium.

\end{appendix}

\bibliographystyle{mn2e}
\bibliography{Lit}

\end{document}

%% file: Befehle.tex
\newcommand{\Kel}{{\rm K}}

\newcommand{\cm}{{\rm cm}}

\newcommand{\GHz}{{\rm GHz}}

\newcommand{\expf}[1]{{{\rm e}^{#1}}}

\newcommand{\ion}[2]{{\text{{\sc #1}\,{\sc #2}}}}

\newcommand{\HeIlevel}[4]{{#1^{#2} {\rm #3}_{#4}}}   

\newcommand{\id}{{\,\rm d}}

\newcommand{\beq}{\begin{equation}}   %

\newcommand{\eeq}{\end{equation}}   %

\newcommand{\beqa}{\begin{eqnarray}}   %

\newcommand{\eeqa}{\end{eqnarray}}   %

\newcommand{\beal}{\begin{align}}
\newcommand{\enal}{\end{align}}

\newcommand{\bspl}{\begin{split}}

\newcommand{\espl}{\end{split}}

\newcommand{\bsub}{\begin{subequations}}

\newcommand{\esub}{\end{subequations}}

\newcommand{\bmulti}{\begin{multline}}   %

\newcommand{\beqm}{\begin{mathletters}}   %

\newcommand{\eeqm}{\end{mathletters}}   %

\newcommand{\Abst}[1]{\,#1}

\newcommand{\me}{m_{\rm e}}

\newcommand{\Ne}{N_{\rm e}}

\newcommand{\Te}{T_{\rm e}}

\newcommand{\Tg}{T_{\gamma}}

\newcommand{\sigT}{\sigma_{\rm T}}

\newcommand{\pot}[2]{#1 \times 10^{#2}}

